\documentclass[twocolumn,showpacs,preprintnumbers,amsmath,amssymb]{revtex4}

\usepackage{graphicx}
\usepackage{dcolumn}
\usepackage{bm}
\def\jpsi{{J/\psi}}

\def\be{\begin{equation}}
\def\ee{\end{equation}}
\def\bea{\begin{eqnarray}}
\def\eea{\end{eqnarray}}
\def\NO{\nonumber}
\def\gev{\mathrm{~GeV}}

\def\dfrac{\displaystyle\frac}
\def\dg{\sp\dagger}
\def\md{\mathrm{d}}
\def\li{\mathrm{Li_2}}
\def\co{{\cal O}}
\def\a{\alpha}
\def\b{\beta}

\def\d{\delta}
\def\e{\epsilon}
\def\g{\gamma}
\def\s{\sigma}
\def\q{\theta}

\def\mopsi{{\langle\mathcal{O}^{\psi}_n\rangle}}
\def\mopsip{{\langle\mathcal{O}^{\psi^\prime}_n\rangle}}

\begin{document}


\title{QCD corrections to polarization of $\jpsi$ and $\Upsilon$ at Tevatron and LHC}

\author{Bin Gong and Jian-Xiong Wang}%
\affiliation{
Institute of High Energy Physics, Chinese Academy of Sciences, P.O. Box 918(4),
Beijing, 100049, China.\\
Theoretical Physics Center for Science Facilities, Beijing, 100049, China.
}%
\date{\today}

\begin{abstract}
In this work, we present more details of the calculation on the next
to leading order (NLO) QCD corrections to polarization of direct
$\jpsi$ production via color singlet at Tevatron and LHC, together
with the results for $\Upsilon$ for the first time. Our results show
that the $\jpsi$ polarization status drastically changes from
transverse polarization dominant at leading order (LO) into
longitudinal polarization dominant in the whole range of the
transverse momentum $p_t$ of $\jpsi$ when the NLO corrections are
counted. For $\Upsilon$ production, the $p_t$ distribution of the
polarization status behaves almost the same as that for $\jpsi$
except that the NLO result is transverse polarization at small $p_t$
range. Although the theoretical evaluation predicts a larger
longitudinal polarization than the measured value at Tevatron, it
may provide a solution towards the previous large discrepancy for
$\jpsi$ and $\Upsilon$ polarization between theoretical prediction
and experimental measurement, and suggests that the next important
step is to calculate the NLO corrections to hadronproduction of
color octet state $\jpsi^{(8)}$ and $\Upsilon^{(8)}$. Our
calculations are performed in two ways, namely we do and do not
analytically sum over the polarizations, and then check them with
each other.
\end{abstract}

\maketitle

\section{Introduction}
The study of $J/\psi$ production on various experiments is a very
interesting topic since its discovery in 1974. It is a good place to
probe both perturbative and nonperturbative aspects of QCD dynamics.
To describe the huge discrepancy of the high-$p_t$ $J/\psi$
production between the theoretical calculation based on color
singlet mechanism\cite{j.h.kuhn:79} and the experimental measurement
by the CDF collaboration at the Tevatron\cite{cdf1}, color-octet
mechanism\cite{Braaten:1994vv} was proposed based on the
non-relativistic QCD(NRQCD)\cite{Bodwin:1994jh}. The factorization
formalism of NRQCD provides a theoretical framework to the treatment
of heavy-quarkonium production. It allows consistent theoretical
prediction to be made and to be improved systematically in the QCD
coupling constant $\a_s$ and the heavy-quark relative velocity $v$.
The color singlet mechanism is straightforward from the perturbative
QCD, but the color-octet mechanism depends on  nonperturbative
universal NRQCD  matrix elements. So various efforts have been made
to confirm this mechanism, or to fix the magnitudes of the universal
NRQCD matrix elements. Although it seems to show qualitative
agreements with experimental data, there are certain difficulties in
the quantitative estimate in NRQCD for $J/\psi$ photoproduction at
the DESY ep collider HERA
\cite{cacciari:1996dg,Amundson:1996ik,Ko:1996xw,Kniehl:1997fv,kramer:1995nb,Kramer:1994zi},
$J/\psi(\psi')$ and $\Upsilon$ polarization of hadronproduction at
the Fermilab Tevatron, and $\jpsi$ production in B-factories. A
review of the situation could be found in Ref.~\cite{kramer:2001}.

Without NLO corrections, it is difficult to obtain agreement between the
experimental results and leading order theoretical predictions for $J/\psi$ production.
There are a few examples shown that NLO corrections are quite large.
It was found that the current experimental results on
inelastic $J/\psi$ photoproduction\cite{hera:h1,hera:zeus} are adequately
described by the color singlet channel alone once higher-order QCD corrections
are included\cite{kramer:1995nb,Kramer:1994zi}.
Although ref.~\cite{Klasen:2001cu} found that the DELPHI
\cite{deBoer:2003xm} data evidently favor the NRQCD formalism for
$J/\psi$ production $\gamma + \gamma \rightarrow J/\psi + X$, rather than the
color-singlet model. And it was also found in ref.~\cite{Qiao:2003ba} that
the QCD higher order process $\gamma + \gamma \rightarrow J/\psi + c + \bar{c}$
gives the same order and even larger contribution at high $p_t$ than
the leading order color singlet processes.
In ref.~\cite{Hagiwara:2007bq}, the NLO process $c + g \rightarrow J/\psi + c$ where the initial $c$ quark is the intrinsic c quark from proton at Tevatron, gives larger contribution at high $p_t$ than the leading order color singlet processes.
The large discrepancies found in the single and double charmonium production in
$e^+e^-$ annihilation at B factories between LO theoretical predictions
\cite{Braaten:2002fi, Liu:2002wq, Hagiwara:2003cw} and experimental results
~\cite{Abe:2002rb,Aubert:2005tj} were studied in many work.
It seems that they may be resolved by including higher order correction: NLO QCD
and relativistic corrections
\cite{Braaten:2002fi, Zhang:2005ch, jxwang:2007je,Gong:2008ce, He:2007te, Zhang:2006ay, Zhang:2008gp}.

Based on NRQCD, the LO calculation predicts a sizable transverse
polarization for $\jpsi$ production at high $p_t$ at
Tevatron\cite{beneke:96yr,braaten:99yr,Leibovich:1996pa} while the
measurement at Fermilab Tevatron \cite{Abulencia:2007us} gives
slight longitudinal polarized result. In a recent paper
\cite{Abazov:2008za}, the measurement on polarization of $\Upsilon$
production at Tevatron is presented and the NRQCD predication
\cite{Braaten:2000gw} is not coincide with it. Beyond the NRQCD
framework, there is a try by using s-channel treatment to $\jpsi$
hadronproduction in the work of ref~\cite{Haberzettl:2007kj}, which
gives longitudinal polarization. Within the NRQCD framework, to
calculate higher order corrections is an important step towards the
solution of such puzzles. Recently, NLO QCD corrections to $\jpsi$
hadronproduction have been calculated  in
ref~\cite{Campbell:2007ws}. The results show that the total cross
section is boosted by a factor of about 2 and the $\jpsi$ transverse
momentum $p_t$ distribution is enhanced more and more as $p_t$
becomes larger. A real correction process $g+g\rightarrow \jpsi
+c+\overline{c}$ at NLO, which is not included in the
ref.~\cite{Campbell:2007ws}, was calculated in
\cite{Qiao:2003ba,Artoisenet:2007xi}. It gives sizable contribution
to $p_t$ distribution of $\jpsi$ at high $p_t$ region, and it alone
gives almost unpolarized result. Therefore it is very interesting to
know the result of $\jpsi$ polarization when NLO QCD corrections are
included. In a recent Letter~\cite{Gong:2008sn}, we presented a
calculation on the NLO QCD corrections to the $\jpsi$ polarization
in hadronproduction at Tevatron and LHC. In this paper, we give more
details of the calculation, and the results for $\Upsilon$
polarization for the first time. The results show that the
polarizations of $\jpsi$ and $\Upsilon$ are drastically changed from
more transverse polarization at LO into more longitudinal
polarization at NLO. Meanwhile, our results for total cross section
and transverse momentum distribution is consistent with
ref.~\cite{Campbell:2007ws}. In this calculation, we use our Feynman
Diagram Calculation package (FDC)\cite{FDC} with newly added part of
a complete set of method to calculate tensor and scalar integrals in
dimensional regularization, which was used in our previous
work\cite{jxwang:2007je,Gong:2008ce}.

This paper is organized as follows. In Sec. II, we give the LO cross section for the process.
The calculation of NLO QCD corrections are described in Sec. III. In Sec. IV, it presents the formula in
final integration to obtain the transverse momentum distribution of $\jpsi$ production.
Sec. V. is devoted to the description about the calculation of $\jpsi$ polarization.
The color factor treatment for all the calculated processes are given in Sec. VI.
In Sec. VII, treatment of $\Upsilon$ is given. The numerical results are presented in Sec. VIII.
Finally, The conclusion and discussion are given in Sec. IX.

\section{The LO cross section of $\jpsi$ hadronproduction}
\begin{figure}
\center{
\includegraphics*[scale=0.6]{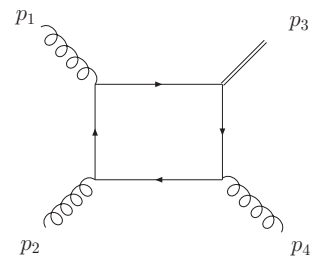}
\caption {\label{fig:LO}Leading order Feynman diagrams for
$g + g \rightarrow \jpsi + g$. The other five diagrams can be obtained by
permutation the places of gluons.}}
\end{figure}

The related Feynman diagrams which contribute to the LO amplitude of the
partonic process $g(p_1)+ g(p_2) \rightarrow \jpsi(p_3) + g(p_4)$ are shown
in Fig.~\ref{fig:LO}, while the others can be obtained by permuting the
places of gluons.

In the nonrelativistic limit, we can use the NRQCD factorization formalism to obtain the partonic differential cross section in $n=4-2\e$ dimension as
\begin{widetext}
\be
\dfrac{\mathrm{d}\hat{\s}^{B}}{\mathrm{d}\hat{t}}=
\dfrac{5\pi\a_s^3|R_s(0)|^2[\hat{s}^2(\hat{s}-1)^2+\hat{t}^2(\hat{t}-1)^2+\hat{u}^2(\hat{u}-1)^2]} {144m_c^5\hat{s}^2(\hat{s}-1)^2(\hat{t}-1)^2(\hat{u}-1)^2} + \co(\e),
\ee
\end{widetext}
by introducing three dimensionless kinematic variables:
\be
\hat{s}=\dfrac{(p_1+p_2)^2}{4m_c^2},\quad \hat{t}=\dfrac{(p_1-p_3)^2}{4m_c^2}, \quad \hat{u}=\dfrac{(p_1-p_4)^2}{4m_c^2},
\ee
where $R_s(0)$ is the radial wave function at the origin of $\jpsi$ and the  reasonable approximation $M_{\jpsi}=2m_c$ is taken.

The LO total cross section is obtained by convoluting the partonic cross section
with the parton distribution function (PDF) $G_g(x,\mu_f)$ in the proton:
\be
\s^B=\int \mathrm{d}x_1\mathrm{d}x_2 G_g(x_1,\mu_f)G_g(x_2,\mu_f)\hat{\s}^B ,
\ee
where $\mu_f$ is the factorization scale. In the following $\hat{\s}$ represents the corresponding partonic cross section.
\section{The NLO cross section of $\jpsi$ hadronproduction}
The NLO contributions to the process can be written as a sum of two
parts: one is the virtual correction which arises from loop
diagrams, the other is the real correction caused by radiation of a
real gluon, or a gluon splitting into a light quark-antiquark pair,
or a light (anti)quark splitting into a light (anti) quark and a
gluon.

\subsection{Virtual corrections}
\begin{figure}
\center{
\includegraphics*[scale=0.40]{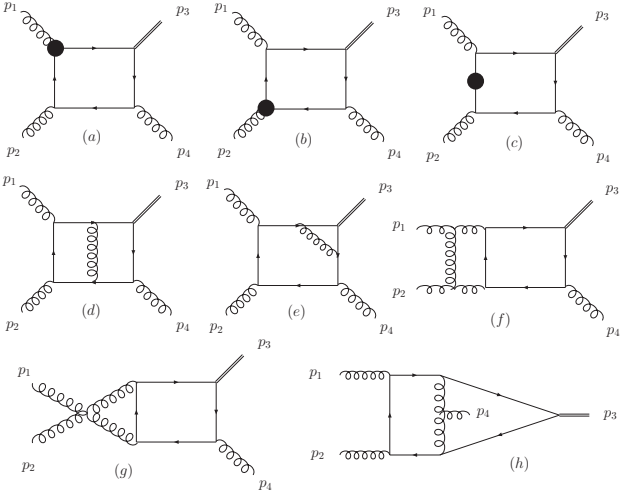}
\caption {\label{fig:NLO}
One-loop diagrams for $g g \rightarrow \jpsi g$.
Group (a) and (b) are
counter-term diagrams of the quark-gluon vertex and corresponding loop diagrams,
Group (c) are the quark self-energy diagrams and corresponding
counter-term ones. More diagrams can be obtained by permutation of
external gluons. }}
\end{figure}
There are UV, IR and Coulomb singularities in the calculation of the virtual
corrections. UV-divergences existing in the self-energy and triangle diagrams are removed by the renormalization of the QCD gauge coupling constant, the charm quark
mass, charm quark and gluon fields. Here we adopt renormalization scheme
used in ref.~\cite{Klasen:2004tz}. For the charm quark mass, charm quark and gluon fields, the renormalization constant $Z_m$, $Z_2$ and $Z_3$ are determined in the on-mass-shell(OS) scheme while for the QCD gauge coupling constant, $Z_g$ is fixed in the modified-minimal-subtraction($\overline{\mathrm{MS}}$) scheme:
\bea
\delta Z_m^{OS}&=&-3C_F\dfrac{\alpha_s}{4\pi}\left[\dfrac{1}{\e_{UV}} -\gamma_E +\ln\dfrac{4\pi \mu_r^2}{m_c^2} +\frac{4}{3} \right] , \NO\\
\delta Z_2^{OS}&=&-C_F\dfrac{\alpha_s}{4\pi}
\left[\dfrac{1}{\e_{UV}} +\dfrac{2}{\e_{IR}} -3\gamma_E +3\ln\dfrac{4\pi \mu_r^2}{m_c^2} +4 \right] , \NO\\
\delta Z_3^{OS}&=&\dfrac{\alpha_s}{4\pi}\left[(\beta_0-2C_A)\left(\dfrac{1}{\e_{UV}} -\dfrac{1}{\e_{IR}}\right)\right] \\
\delta Z_g^{\overline{\mathrm{MS}}}&=&-\dfrac{\beta_0}{2}\dfrac{\alpha_s}{4\pi}\left[\dfrac{1}{\e_{UV}} -\gamma_E +\ln(4\pi) \right] . \NO
\eea
where $\g_E$ is Euler's constant, $\b_0=\frac{11}{3}C_A-\frac{4}{3}T_Fn_f$ is the
one-loop coefficient of the QCD beta function and $n_f$ is the number of active
quark flavors. There are three massless light quarks $u, d, s$, so $n_f$=3. In $SU(3)_c$, color factors are given by
$T_F=\frac{1}{2}, C_F=\frac{4}{3}, C_A=3$. And $\mu_r$ is the renormalization scale.

After having fixed the renormalization scheme, there are 129 NLO diagrams in total, including counter-term diagrams. They are shown in Fig.~\ref{fig:NLO}, and divided into 8 groups. Diagrams of group
$(e)$ that has a virtual gluon line connected with the quark pair lead to
Coulomb singularity, which can be isolated by introducing
a small relative velocity $v=|\vec{p}_{c}-\vec{p}_{\bar{c}}|$. The corresponding
contribution is also of $\co(\a_s)$ and can be mapped into the $c\bar{c}$
wave function.
\bea
\s&=&|R_s(0)|^2\hat{\s}^{(0)}\left( 1 +\dfrac{\a_s}{\pi}C_F\dfrac{\pi^2}{v} + \dfrac{\a_s}{\pi}C +\co(\a_s^2)\right)  \NO\\
&\Rightarrow&|R^{ren}_s(0)|^2 \hat{\s}^{(0)} \left[1+ \dfrac{\a_s}{\pi}C +\co(\a_s^2)\right] .
\eea

The Passarino-Veltman reduction \cite{Passarino:1978jh} is adopted in
the tensor decomposition when it's Gram determinant
is nonzero. Otherwise, It is to do the
integration directly with Feynman parametrization for two-point tensor case, 
and to write the Lorentz structure with
independent external momentums and apply the Passarino-Veltman
reduction again for other cases.
In the calculation of scalar integral, we first
try to decompose the scalar integral into several lower-point ones when it's Gram determinant
is zero, if it fails, then to do the integration directly with Feynman parametrization 
just like the treatment to scalar integral with nonzero Gram determinants. 
Above procedure, including both reduction and integration, are done by FDC automatically.

In our calculation, there are total 86 scalar integrals in total:
\begin{itemize}
\item 65 of the total 86 integrals, can be found in Ref.~\cite{kramer:1995nb} after including the
permutation of s, t and u. But the explicit results for the three Coulomb
singular five-point scalar integrals is not available in Ref.~\cite{kramer:1995nb}.
\item The remaining 21 integrals are not listed in Ref.~\cite{kramer:1995nb}. 12 of them can be reduced to combination of some lower-point scalar integrals
and needn't to be integrated directly.
\item Another 6 of them can be expressed by the following two integrals,
$C(p_1,p_3,m_c,m_c,m_c)$
and 
$D(p_1,p_4,p_3+p_4,0,m_c,m_c,m_c)$,
through permutation of s, t and u, where $A, B, C, D, E$ are defined exactly the same as in Ref.~\cite{kramer:1995nb}. They can be written into a linear
combination of another two scalar integrals as: 
\bea
&&C(p_1,p_3,m_c,m_c,m_c)\NO\\
&=&\dfrac{1}{2}C(-p_3/2,-p_3/2+p_1,0,m_c,m_c) \NO\\
&&+\dfrac{1}{2}C(p_3/2,-p_3/2+p_1,0,m_c,m_c), \\
&&D(p_1,p_4,p_3+p_4,0,m_c,m_c,m_c) \NO\\
&&=\dfrac{1}{2}D(p_3/2,p_3/2-p_2,-p_3/2-p_4,m_c,m_c,m_c,m_c)\NO\\
&+&\dfrac{1}{2}D(-p_3/2,p_3/2-p_2,-p_3/2-p_4,m_c,m_c,m_c,m_c). \NO
\eea
But in our calculation, they are calculated independently, and above
relationship can be used to check all three scalar integrals.
\item The remaining 3 scalar integrals can be expressed by $B(p_1,m_c,m_c)$ through the permutation of s, t and u.
\end{itemize} 
More details about these 86 scalar integrals can be found at FDC homepage\footnote{
http://www.ihep.ac.cn/lunwen/wjx/public\_html/2008 /gggjpsi/index.html}.

By adding all diagrams together, the virtual corrections to the
differential cross section can be expressed as
\be
\dfrac{\mathrm{d}\hat{\s}^{V}}{\mathrm{d}t} \propto 2\mathrm{Re}(M^BM^{V*}),
\ee
where $M^B$ is the amplitude at LO, and $M^V$ is the renormalized
amplitude at NLO. $M^V$ is UV and Coulomb finite,
but it still contains the IR divergences:
\be
M^V|_{IR}=\left[ \frac{\a_s}{2\pi} \frac{\Gamma(1-\e)}{\Gamma(1-2\e)} \left(\frac{4 \pi \mu_r^2}{s_{12}}\right)^{\e}\right]\left(\dfrac{A^V_2}{\e^2} + \dfrac{A^V_1}{\e}\right) M^B,
\ee
with
\be
A^V_2=-\dfrac{9}{2}, \quad A^V_1=-\dfrac{3}{2}\biggl[ \ln\biggl(\dfrac{\hat{s}}{-\hat{t}}\biggr) +\ln\biggl(\dfrac{\hat{s}}{-\hat{u}}\biggr) \biggr] +\dfrac{1}{2}n_f -\dfrac{33}{4} .
\ee
And the total cross section of virtual contribution could be written as:
\be
\s^V=\int \mathrm{d}x_1\mathrm{d}x_2 G_g(x_1,\mu_f)G_g(x_2,\mu_f)\hat{\s}^V .
\ee
\subsection{Real corrections}
The real corrections arise from four parton level subprocesses:
\bea
g(p_1) + g(p_2) &\rightarrow& \jpsi (p_3) + g(p_4) + g(p_5) ,\label{prs:gggg}\\
g(p_1) + g(p_2) &\rightarrow& \jpsi (p_3) + q(p_4) + \overline{q}(p_5) ,\label{prs:ggqq}\\
g(p_1) +q(\overline{q})(p_2) &\rightarrow& \jpsi (p_3) + g(p_4) +
q(\overline{q})(p_5) .\label{prs:qggq} \\
g(p_1) + g(p_2) &\rightarrow& \jpsi (p_3) + c(p_4) + \overline{c}(p_5) ,\label{prs:ggcc}
\eea

\begin{figure}
\center{
\includegraphics*[scale=0.5]{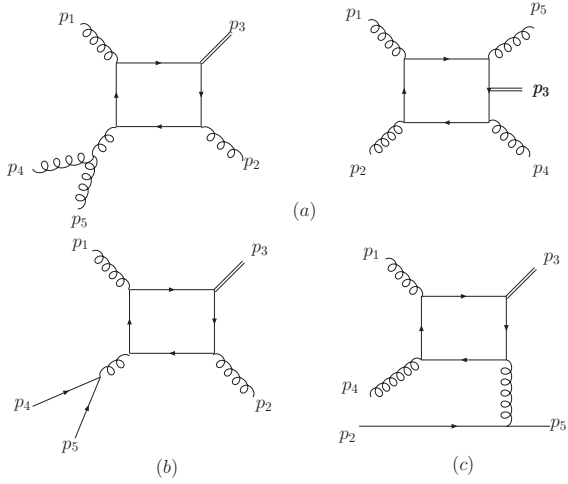}
\caption {\label{fig:real}Feynman diagrams for first three real correction processes. (a) is
for $gg \rightarrow \jpsi + gg$ and (b) is for
$gg \rightarrow \jpsi + q\overline{q}$ while (c) is for
$gq(\overline{q}) \rightarrow \jpsi + gq(\overline{q})$.
More diagrams can be obtained by all possible permutation of gluons.}}
\end{figure}

\begin{figure}
\center{
\includegraphics*[scale=0.5]{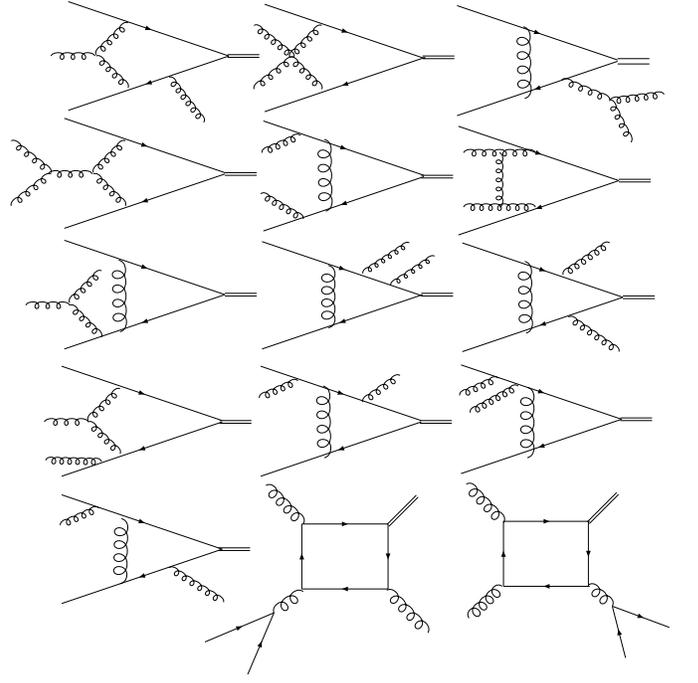}
\caption {\label{fig:ggcc}Feynman diagrams for real correction process $gg \rightarrow \jpsi + c\overline{c}$. More diagrams can be obtained by reversing the arrow of charm quark line and/or exchanging the places of gluons.}}
\end{figure}

We have neglected the contribution from a real correction subprocess
$q\bar{q}\rightarrow \jpsi gg$, which is IR finite and tiny (it only
contributes about 0.002\% at $p_t$=3 GeV and 0.05\% at $p_t$=50 GeV
to the differential cross section). And Feynman diagrams for above processes are shown in Fig.~\ref{fig:real} and Fig.~\ref{fig:ggcc}.
The phase space integration of above processes (except $gg \rightarrow \jpsi + c\overline{c}$) generates IR singularities,
which are either soft or collinear and can be conveniently isolated
by slicing the phase space into different regions. We use the two-cutoff
phase space slicing method \cite{Harris:2001sx} to introduces two small
cutoffs to decompose the phase space into three parts.

Real gluon emission brings soft singularities. A small soft cutoff $\d_s$ is
used to divide the phase space into two regions according to that the emitted
gluon is soft or hard. Then another small cutoff $\d_c$ is used to decompose
the hard region into collinear and noncollinear regions.
Then the cross section of real correction processes can be written as
\be
\s^R=\s^S+\s^{HC}+\s^{H\overline{C}} .
\ee
The hard noncollinear part $\s^{H\overline{C}}$ is IR finite
and can be numerically computed using standard Monte-Carlo integration techniques. The subprocess $gg \rightarrow \jpsi + c\overline{c}$ consists of only  hard noncollinear part.
\subsubsection{soft}
It is easy to find that soft singularities caused by emitting soft gluons from the
charm quark-antiquark pair in the S-wave color singlet $\jpsi$ are canceled
by each other.
Therefore only the real gluon emission subprocess in Eq.~(\ref{prs:gggg}), where
there could be a soft gluon emitted from the external gluons, contains soft
singularities. Suppose $p_5$ is the momentum of the emitted gluon. If we define the
Mandelstam invariants as $s_{ij}=(p_i+p_j)^2$ and $t_{ij}=(p_i-p_j)^2$,
the soft region is defined in term of the energy of $p_5$ in the $p_1+p_2$ rest frame
by $0 \leq E_5 \leq \d_s \sqrt{s_{12}}/2$. In this region,
soft approximation can be made and the matrix element squared can be
factorized as
\be
|M_R|^2|_{\rm soft} \simeq -4\pi\a_s \mu_r^{2\e} \sum_{i,j=1,2,4}
\frac{-p_i \cdot p_j}{(p_i \cdot p_5)(p_j\cdot p_5)} M^0_{ij} \, ,
\label{eqn:sme_soft}
\ee
with
\be
M^0_{ij} =
\left[{\bf T}^a(i) {\bf M}_{b_1\cdots b_{i^\prime}\cdots b_4}^B\right]^{\dg} \left[{\bf T}^a(j)
{\bf M}_{b_1\cdots b_{j^\prime}\cdots b_4}^B \right] \,
\label{eqn:me0_cc}
\ee
and
\be
{\bf T}^a(j)=if_{ab_jb_{j^\prime}},
\ee
where ${\bf M}_{b_1\cdots b_4}^B$ is the color connected Born matrix element.

Meanwhile, if we parametrize the emitted gluon's $n$-dimension momentum in
the $p_1+p_2$ rest frame as
\be
p_5=E_5(1, \ldots, \sin\q_1 \cos\q_2, \cos\q_1) \, ,
\label{eqn:p5_soft}
\ee
the three-body phase space in the soft limit can also be factorized as
\be
\md\Gamma_3|_{\rm soft} = \md\Gamma_2 \left[ \left( \frac{4\pi}{s_{12}} \right)^\e
         \frac{\Gamma(1-\e)}{\Gamma(1-2\e)} \frac{1}{2(2\pi)^2} \right] \md S \, ,
\label{eqn:soft_PS3}
\ee
with
\bea
\md S &=&  \frac{1}{\pi}
     \left( \frac{4}{s_{12}} \right)^{-\e} \int_0^{\d_s\sqrt{s_{12}}/2}
     \md E_5 E_5^{1-2\e} \sin^{1-2\e}\!\q_1\,d\q_1
\NO\\&&\times
\sin^{-2\e}\!\q_2\,d\q_2 \, ,
\label{eqn:soft_PS}
\eea
as given in ref.~\cite{Harris:2001sx}.
After analytical integration over the soft gluon phase space, the parton
level cross section in soft region can be expressed
\be
\hat{\s}^S=\hat{\s}^B\left[ \frac{\a_s}{2\pi} \frac{\Gamma(1-\e)}{\Gamma(1-2\e)} \left(\frac{4 \pi \mu_r^2}{s_{12}}\right)^{\e}\right]\left(\dfrac{A^S_2}{\e^2} + \dfrac{A^S_1}{\e} + A^S_0\right)
\ee
with
\be
A^S_2=9, \quad A^S_1=3\biggr[\ln\biggl(\dfrac{\hat{s}-1}{-\hat{t}}\biggr) + \ln\biggl(\dfrac{\hat{s}-1}{-\hat{u}}\biggr)\biggr] -18\ln\d_s ,
\ee
and
\bea
A^S_0&=&
18\ln^2\d_s
-6\ln\d_s\biggl[\ln\biggl(\dfrac{\hat{s}-1}{-\hat{t}}\biggr) + \ln\biggl(\dfrac{\hat{s}-1}{-\hat{u}}\biggr)\biggr]
\NO\\&&
+\dfrac{3}{2}\biggl[\ln^2\biggl(\dfrac{\hat{s}-1}{-\hat{t}}\biggr) +\ln^2\biggl(\dfrac{\hat{s}-1}{-\hat{u}}\biggr)\biggr]
 \NO\\&&
+ 3\biggl[\li\biggl(\dfrac{-\hat{t}}{\hat{s}-1}\biggr) + \li\biggl(\dfrac{-\hat{u}}{\hat{s}-1}\biggr)\biggr].
\eea
\subsubsection{hard collinear}
The hard collinear regions of the phase space are those where any invariant
($s_{ij}$ or $t_{ij}$) becomes smaller in magnitude than $\d_c s_{12}$.
It is treated according to whether the singularities are from initial or final
state emitting or splitting in the origin. Subprocess in Eq.~(\ref{prs:ggqq}) contains final state collinear singularities, and subprocess in Eq.~(\ref{prs:qggq}) contains initial
state collinear singularities while subprocess in Eq.~(\ref{prs:gggg}) contains both.
\paragraph{final state collinear}
For subprocesses in Eq.~(\ref{prs:gggg}) and (\ref{prs:ggqq}), the final state collinear
region is defined by $0 \le s_{45} \le \d_c s_{12}$. As a consequence of the
factorization derivation \cite{Collins:1985ue,Bodwin:1984hc}, the squared
matrix element factorizes into the product of a splitting kernel
and the LO squared matrix element.
\be
|M_R|^2|_{\rm coll} \simeq 4\pi\a_s \mu_r^{2 \e} \frac{2}{s_{45}}P_{44'}(z,\e) |M^B|^2  \, ,
\ee
where $4'$ denotes the parton which splits into parton 4 and 5 collinear pair and
$P_{ij}(z,\e)$ are the unregulated ($z<1$) splitting functions in $n=4-2\e$
dimensions related to the usual Altarelli-Parisi splitting kernels
\cite{Altarelli:1977zs} with $z$ denoting the fraction of the momentum
of parton $4'$ carried by parton $4$. For $z<1$ the $n-$dimensional
unregulated splitting functions are written as
$P_{ij}(z,\e)=P_{ij}(z)+\e P^{\prime}_{ij}(z) $ with
\bea
P_{qq}(z) &=& C_F \frac{1+z^2}{1-z} ,\nonumber \\
P_{qq}^{\prime}(z) &=& -C_F(1-z) ,\nonumber\\
P_{gg}(z) &=& 6\left[ \frac{z}{1-z}+\frac{1-z}{z}+z(1-z)\right] ,\nonumber\\
P_{gg}^{\prime}(z) &=& 0 ,\nonumber\\
P_{qg}(z) &=& \frac{1}{2} \left[ z^2+(1-z)^2 \right] ,\nonumber\\
P_{qg}^{\prime}(z) &=& -z(1-z) \, .
\eea
Meanwhile, the three-body phase space in the collinear limit can also be
factorized as \cite{Harris:2001sx}:
\be
\md\Gamma_3|_{\rm coll} = \md\Gamma_2 \frac{(4\pi)^\e}{16\pi^2\Gamma(1-\e)} \md z \md s_{45} [s_{45}z(1-z)]^{-\e} \, .
\ee
Hence after integrations of $z$ and $s_{45}$, the parton level cross section
in hard final state collinear region can be expressed as
\bea
\hat{\s}^{HC}_f&=&\hat{\s}^B\left[ \frac{\a_s}{2\pi} \frac{\Gamma(1-\e)}{\Gamma(1-2\e)} \left( \frac{4\pi\mu_r^2}{s_{12}} \right)^\e \right]
\\&&
\times\biggl( \frac{A_1^{g \rightarrow gg}+A_1^{g \rightarrow q\overline{q}}}{\e} + A_0^{g \rightarrow gg}+A_0^{g \rightarrow q\overline{q}} \biggr) \, .\NO
\eea
where $A_1$ and $A_0$ are
\bea
A_1^{g \rightarrow gg} &=& 3 \left( 11/6 + 2 \ln\d'_s \right) \nonumber\\
A_0^{g \rightarrow gg} &=& 3 \left[ 67/18 - \pi^2/3 - \ln^2\d'_s
    - \ln\d_c \left( 11/6 + 2 \ln\d'_s \right) \right] \nonumber\\
A_1^{g \rightarrow q\overline{q}} &=& -n_f/3  \nonumber\\
A_0^{g \rightarrow q\overline{q}} &=& n_f/3 \left( \ln\d_c-5/3 \right) \, ,
\eea
for subprocesses in Eq.~(\ref{prs:gggg}) and (\ref{prs:ggqq}), and
\be
\d'_s = \frac{s_{12}}{s_{12}+s_{45}-M^2_{\jpsi}} \simeq \frac{\hat{s}}{\hat{s}-1} \d_s \, .
\ee
Thus the total cross section for real correction processes in hard final state
collinear region can be written as:
\be
\s^{HC}_f=\int \mathrm{d}x_1\mathrm{d}x_2 G_g(x_1,\mu_f)G_g(x_2,\mu_f)\hat{\s}^{HC}_f .
\ee
\paragraph{initial state collinear}
For subprocess in Eq.~(\ref{prs:qggq}), the hard initial state collinear region is
defined by $0 \le -t_{25} \le \d_c s_{12}$. However for subprocess in Eq.~(\ref{prs:gggg}),
the hard initial state collinear region is defined if any of the following
conditions is satisfied $0 \le -t_{ij} \le \d_c s_{12}$, with $i=1,2$
and $j=3,4$. For convenience, suppose that $2$ and $5$ are the partons involved
in the splitting $2 \rightarrow 2'+5$ while $2'$ denotes an internal gluon.
Following the similar way as in the final state collinear case,
the squared matrix element can be written as
\be
|M_R|^2|_{\rm coll} \simeq  4\pi\a_s \mu_r^{2 \e} \frac{2}{-zt_{25}} P_{2'2}(z,\e)  |M^B|^2 \, ,
\ee
where $z$ denotes the fraction of parton 2's momentum carried by parton $2'$
with parton 5 taking a fraction $(1-z)$. And the three-body phase space in
the collinear limit can also be factorized as:
\be
\md\Gamma_3|_{\rm coll} = \md\Gamma_2 \frac{(4\pi)^\e}{16\pi^2\Gamma(1-\e)} \md z \md t_{25} [-(1-z)t_{25}]^{-\e} \, .
\ee
The $t_{25}$ integration yields
\be
\int_0^{\d_c s_{12}} -\md t_{25} (-t_{25})^{-1-\e} = -\frac{1}{\e} (\d_c s_{12})^{-\e} \, .
\ee
If we write the total cross section of LO as
\be
\md \s^B=\md x_1\md x_2 G_g(x_1)G_g(x_2)\md \hat{\s}^B ,
\label{eqn:ds_LO}
\ee
where $G_g(x_i)$ is the bare PDF. And using above results, the three-body
cross section in the hard initial state collinear region can be written
as \cite{Harris:2001sx}
\bea
\md\s^{HC}_i & = & G_g(x_1)G_2(y)\md y \md\hat{\s}^B (zs_{12},t_{13},t_{14})
\NO\\&&\times
\left[ \frac{\a_s}{2\pi} \frac{\Gamma(1-\e)}{\Gamma(1-2\e)}
\left(\frac{4 \pi \mu_r^2}{s_{12}}\right)^{\e}\right]
\NO\\&&\times
\left(-\frac{1}{\e}\right)
\d_c^{-\e}P_{2'2}(z,\e)\md z (1-z)^{-\e}
\NO\\&&\times
\d(yz-x_2)\md x_1\md x_2 \, .
\eea
Notice that a factor of $1/z$ has been absorbed into the flux factor for the
two-body subprocess, and the delta function used here ensures that the fraction
of hadron's momentum carried by $2'$ is $x_2$. And one more thing that
need to be cared is, $s_{12}$ here is related to the square of the overall
hadronic squared center-of-mass energy $S$ by $s_{12}=x_1yS$, but in the
LO process the relation is $s_{12}=x_1x_2S$. From now on, we take the latter
definition, so that the replacement $s_{12} \rightarrow ys_{12}/x_2$
should be made. After the $y$ integration we have
\bea
\md\s^{HC}_i
& = & G_g(x_1)G_2(x_2/z) \md\hat\s^B\left[ \frac{\a_s}{2\pi} \frac{\Gamma(1-\e)}{\Gamma(1-2\e)} \left(\frac{4\pi\mu_r^2}{s_{12}}\right)^{\e}\right] \nonumber \\
&\times& \left(-\frac{1}{\e}\right) \d_c^{-\e}P_{2'2}(z,\e)\frac{dz}{z} \left[ \frac {(1-z)}{z} \right]^{-\e}
\md x_1 \md x_2 \, . \NO\\
\label{eqn:ds_coll} \eea When all possible two-to-three subprocesses
are considered, there will be several contributions, corresponding
to a sum over all possible parton 2. It can be $2=g$ followed by
$g\rightarrow gg$ or $2=q(\overline{q})$ followed by
$q(\overline{q})\rightarrow q(\overline{q})g$. The collinear
singularity must be factorized and absorbed into the redefinition of
the PDF, which is in general called mass factorization
\cite{Altarelli:1979ub}. Here we adopt a scale dependent PDF using
the modified minimal subtraction $(\overline{\rm MS})$ convention
given by \cite{Harris:2001sx}. \bea
G_b(x,\mu_f)&=&G_b(x)+\left(-\frac{1}{\e}\right) \left[
\frac{\a_s}{2\pi} \frac{\Gamma(1-\e)}{\Gamma(1-2\e)}
\left(\frac{4\pi \mu_r^2}{\mu_f^2}\right)^{\e}\right] \NO\\&&\times
\int_x^1 \frac{\md z}{z} P_{bb'}(z)G_{b'}(x/z) \, . \label{eqn:PDF}
\eea Use this definition to replace $G_g(x_2)$ in the LO expression
(\ref{eqn:ds_LO}) and combine the result with the hard initial
collinear contribution (\ref{eqn:ds_coll}), then the resulting
$\co(\a_s)$ expression for the hard initial collinear contribution
is \cite{Harris:2001sx} \bea \md \s^{HC}_i &=& \md\hat\s^B \left[
\frac{\a_s}{2\pi} \frac{\Gamma(1-\e)}{\Gamma(1-2\e)} \left(\frac{4
\pi \mu_r^2}{s_{12}}\right)^{\e}\right] \NO\\&&
 \times \left\{ G_g(x_1,\mu_f)\widetilde{G}_g(x_2,\mu_f)
+ \biggl[ \frac{A_1^{sc}(g\rightarrow gg)}{\e} \right.\\&&\left. +
A_0^{sc} (g\rightarrow gg)\biggr]
G_g(x_1,\mu_f)G_g(x_2,\mu_f)\right\}\md x_1 \md x_2 .\NO
\label{eqn:sig_coll} \eea with \be \widetilde{G}_c(x,\mu_f) =
\sum_{c'}  \int_x^{1-\d_s\d_{cc'}} \frac{dy}{y} G_{c'}(x/y,\mu_f)
\widetilde{P}_{cc'}(y) \, , \label{eqn:g_tilde} \ee and \be
\widetilde{P}_{ij}(y) = P_{ij}(y)\ln\left(\d_c\frac{1-y}{y}
\frac{s_{12}}{\mu_f^2}\right) - P_{ij}^{\prime}(y) \, . \ee The soft
collinear factors $A_i^{sc}$ result from the mismatch in the $z$
integrations. They are given by $A_0^{sc}=A_1^{sc}
\ln(s_{12}/\mu_f^2)$ and $A_1^{sc}(g\rightarrow gg) = 6 \ln \d_s +
(33-2 n_f)/6$. For subprocess in Eq.~(\ref{prs:qggq}), the light
quark(antiquark) can come from either initial hadrons, while for
subprocess in Eq.~(\ref{prs:gggg}), initial collinear may happen to
either of the initial gluons, thus the cross section of hard initial
collinear regions can be written as \be \s^{HC}_i = \s^{HC}_{add} +
\int \hat\s^{HC}_{i} G_g(x_1,\mu_f)G_g(x_2,\mu_f)\md x_1 \md x_2,
\ee with \bea \s^{HC}_{add}&\equiv&\int \hat\s^B \left[
\frac{\a_s}{2\pi} \frac{\Gamma(1-\e)}{\Gamma(1-2\e)} \left(\frac{4
\pi \mu_r^2}{s_{12}}\right)^{\e}\right]
\\&&\times
 \biggl[G_g(x_1,\mu_f)\widetilde{G}_g(x_2,\mu_f) +(x_1\leftrightarrow x_2)\biggr] \md x_1 \md x_2, \NO
\eea
and
\bea
\hat\s^{HC}_{i}&=&2\hat\s^B\left[ \frac{\a_s}{2\pi} \frac{\Gamma(1-\e)}{\Gamma(1-2\e)} \left(\frac{4 \pi \mu_r^2}{s_{12}}\right)^{\e}\right]
\NO\\&&\times
\left[ \frac{A_1^{sc}(g\rightarrow gg)}{\e} + A_0^{sc} (g\rightarrow gg)\right] .
\eea
\subsection{Cross section of all NLO contributions}
The cross section of real correction processes in hard noncollinear regions
could be written as
\begin{widetext}
\bea
\s^{H\overline{C}}&=&\int \biggl[\hat\s^{H\overline{C}}(gg\rightarrow \jpsi +gg) +\sum_{q=u,d,s,c}\hat\s^{H\overline{C}}(gg\rightarrow \jpsi +q\overline{q})
\biggr] \md x_1 \md x_2 G_g(x_1,\mu_f)G_g(x_2,\mu_f) \NO\\
&&+ \int \sum_{\a=u,d,s,\overline{u},\overline{d},\overline{s}}\hat\s^{H\overline{C}}(
g\a\rightarrow \jpsi +g\a)\biggl[G_g(x_1,\mu_f)G_\a(x_2,\mu_f) +(x_1\leftrightarrow x_2)\biggr] \md x_1 \md x_2 ,\NO\\
\eea
\end{widetext}
Thus the cross section of all real corrections becomes
\bea
\s^R&=&\s^{HC}_{add} +\s^{H\overline{C}}
+\int \bigl(\hat\s^S +\hat\s^{HC}_{f} +\hat\s^{HC}_{i}  \bigr)
\NO\\&&\times
G_g(x_1,\mu_f)G_g(x_2,\mu_f)\md x_1 \md x_2 .
\eea
And the total cross section of NLO QCD correction is
\bea
\s^{NLO}=\s^{HC}_{add} +\s^{H\overline{C}} + \s^{V^+},
\eea
with
\bea
\s^{V^+} &\equiv& \int \bigl(\hat\s^B +\hat\s^V +\hat\s^S +\hat\s^{HC}_{f} +\hat\s^{HC}_{i} \bigr)
\NO\\&&\times
G_g(x_1,\mu_f)G_g(x_2,\mu_f)\md x_1 \md x_2 .
\eea
It is easy to find that there is no IR singularities in above expression,
for $2A^V_2+A^S_2=0$ and $2A^V_1+A^S_1+A_1^{g \rightarrow gg}+A_1^{g
\rightarrow q\overline{q}} +2A_1^{sc}(g\rightarrow gg) =0$. The apparent
logarithmic $\d_s$ and $\d_c$ dependent terms also cancel
after numerically integration over the phase space.
\section{Transverse momentum distribution}
To obtain the transverse momentum distribution of $\jpsi$, a transformation for integration variable ($\md x_2 \md t \rightarrow \md p_t \md y$) is introduced. Thus we have
\bea
\s&=&\displaystyle{\int}
\md x_1 \md x_2 \md t G_g(x_1,\mu_f)G_g(x_2,\mu_f)\dfrac{\md \hat \s}{\md t} \NO\\
&=&\int J \md x_1 \md p_t \md y G_g(x_1,\mu_f)G_g(x_2,\mu_f) \dfrac{\md \hat \s}{\md t} ,
\eea
and
\bea
\dfrac{\md \s}{\md p_t}=
\int J \md x_1 \md y G_g(x_1,\mu_f)G_g(x_2,\mu_f) \dfrac{\md \hat \s}{\md t},
\eea
with
\bea
&p_1=x_1\dfrac{\sqrt{S}}{2}(1,0,0,1),
&p_2=x_2\dfrac{\sqrt{S}}{2}(1,0,0,-1), \NO\\[3mm]
&m_t=\sqrt{M_{\jpsi}^2+p_t^2},
&p_3=(m_t \cosh y,p_t,0,m_t \sinh y),\NO\\[3mm]
&x_t=\dfrac{2m_t}{\sqrt{S}},
&\tau=\dfrac{m_4^2-M_{\jpsi}^2}{\sqrt{S}},\\[3mm]
&J=\dfrac{4 x_1 x_2 p_t}{2x_1-x_t e^y},
&x_2=\dfrac{2\tau+x_1~x_t e^{-y}}{2 x_1-x_t e^y}, \NO\\[3mm]
&x_1|_{min}=\dfrac{2 \tau + x_t e^y}{2- x_t e^{-y}},&\NO
\eea
where $\sqrt{S}$ is the center-of-mass energy of $p\bar{p}(p)$ at Tevatron or LHC,
$m_4$ is the invariant mass of all the final state particles except $\jpsi$, and
$y$ and $p_t$ are the rapidity and transverse momentum of $\jpsi$ in the laboratory frame respectively.

\section{Polarization}
The polarization factor $\alpha$ is defined as:
\be
\alpha(p_t)=\frac{{\md\s_T}/{\md p_t}-2 {\md\s_L}/{\md p_t}}
                 {{\md\s_T}/{\md p_t}+2 {\md\s_L}/{\md p_t}}
\ee It represents the measurement of $\jpsi$ polarization as
function of $\jpsi$ transverse momentum $p_t$ when calculated at
each point in $p_t$ distribution. To calculate $\a(p_t)$, the
polarization of $\jpsi$ must be explicitly retained in the
calculation. The partonic differential cross section for a polarized
$\jpsi$ could be expressed as: \be \dfrac{\md
\hat{\s}_{\lambda}}{\md t}= a~\epsilon(\lambda) \cdot
\epsilon^*(\lambda) + \sum_{i,j=1,2} a_{ij} ~p_i \cdot
\epsilon(\lambda) ~p_j \cdot \epsilon^*(\lambda), \ee where
$\lambda=T_1,T_2,L$. $\epsilon(T_1),~\epsilon(T_2),~\epsilon(L)$ are
the two transverse polarization vectors and the longitudinal
polarization one of $\jpsi$, and the polarization of all the other
particles are summed over in n-dimensions. It causes a more
difficult tensor reduction path than that with all the polarizations
being summed over in the calculation of virtual corrections. It is
found that $a$ and $a_{ij}$ are finite when the virtual corrections
and real corrections are summed up. Therefore there is no difference
in the differential cross section ${\md \hat{\s}_{\lambda}}/{\md t}$
whether the polarization of $\jpsi$ is summed over in 4 or $n$
dimensions. Thus we can just treat the polarization vectors of
$\jpsi$ in 4-dimension, and also the spin average factor goes back
to 4-dimension.

To make a cross check, we carry out another calculation. Namely, we calculate the differential cross section $\s^{HC}_{add}$ and $\s^{V^+}$ with the
the polarizations of all particles being summed up analytically. The results are numerically compared with that obtained without summing up the polarization of $\jpsi$.
Moreover, to check gauge invariance, in the expression we explicitly keep the gluon polarization vector and then replace it by its 4-momentum in the final numerical calculation. Definitely the result must be zero and our results confirm it.
To calculate $\s^{H\bar{C}}$, only numerical computation is carried out and we only sum over the physical polarizations of the gluons
to avoid involving diagrams which contain external ghosts lines.

\section{Color Factor}
There is just one color factor $d_{c_1c_2c_4}$ for the LO process in amplitude
level with $c_1,c_2$ and $c_4$ being the color indices of the three gluons in the process. And it is the same for the virtual correction process that just only one color factor
$d_{c_1c_2c_4}$ appears in amplitude level. For other processes, color factors are orthogonalized and normalized. There are three color factors in amplitude level for real correction process $g+g\rightarrow\jpsi+g+g$
\bea
&\dfrac{1}{\sqrt{5}}\mathrm{Tr}\bigl[T^{c_4}T^{c_1}T^{c_5}T^{c_2}-
T^{c_4}T^{c_2}T^{c_5}T^{c_1}\bigr],\NO\\
&\dfrac{1}{\sqrt{5}}\mathrm{Tr}\bigl[T^{c_4}T^{c_5}T^{c_1}T^{c_2}-
T^{c_4}T^{c_2}T^{c_1}T^{c_5}\bigr],\\
&\dfrac{1}{\sqrt{5}}\mathrm{Tr}\bigl[T^{c_4}T^{c_1}T^{c_2}T^{c_5}-
T^{c_4}T^{c_5}T^{c_2}T^{c_1}\bigr],\NO
\eea
where $c_i$ are the color indices of the external gluons.
For $g+g\rightarrow\jpsi+q+{\bar q}$, there is one color factor
\be
\dfrac{\sqrt{3}}{6\sqrt{5}}\bigl[3(T^{c_1}T^{c_2} +T^{c_2}T^{c_1})_{c_4c_5}
- \delta_{c_4c_5}\delta_{c_1c_2}\bigr],
\ee
where $c_1,c_2$ and $c_4,c_5$ are the color indices of the external gluons and quark, anti quarks
respectively. And $g+q\rightarrow\jpsi+g+q$ has almost the same color factor as above.
For $g+g\rightarrow\jpsi+c+{\bar c}$, there are three color factors
\bea
&\dfrac{1}{2\sqrt{66}}\bigl[6(T^{c_2}T^{c_1})_{c_4c_5} + \delta_{c_4c_5}\delta_{c_1c_2}\bigr],\NO\\
&\dfrac{1}{2\sqrt{858}}\bigl[4(T^{c_2}T^{c_1})_{c_4c_5}-22(T^{c_1}T^{c_2})_{c_4c_5}
-3\delta_{c_4c_5}\delta_{c_1c_2}\bigr],\\
&\dfrac{3\sqrt{26}}{52\sqrt{15}}\bigl[4(T^{c_2}T^{c_1})_{c_4c_5}+4(T^{c_1}T^{c_2})_{c_4c_5}
-3\delta_{c_4c_5}\delta_{c_1c_2}\bigr],\NO
\eea
where $c_1,c_2$ and $c_4,c_5$ are the color indices of the external gluons and c quark, anti c quarks respectively.
\section{Treatment of $\Upsilon$}
The production mechanism of $\Upsilon$ at Tevatron and LHC is much
similar to that of $\jpsi$ except that, color octet states
contribute much less in $\Upsilon$ production according to the
experimental data and LO theoretical predictions. We can apply the
results of above calculation to the case of $\Upsilon$ by doing the
substitutions: \bea
m_c&\leftrightarrow& m_b   \NO\\
M_{\jpsi}&\leftrightarrow& M_\Upsilon \NO\\
R_s(0)^{\jpsi}&\leftrightarrow& R_s(0)^\Upsilon \\
n_f=3 &\leftrightarrow& n_f=4  \NO \eea Note that charm quark is
treated as light quark as an approximation. It is not coincide with
the definition of CTEQ6M PDFs used in the calculation. The mass of
heavy quark is not zero in the definition of CTEQ6M PDFs. This
approximation can cause a small uncertainty.

\section{Numerical result}
In our numerical calculations, the CTEQ6L1 and CTEQ6M PDFs \cite{cteq}, and the corresponding fitted value for $\a_s(M_Z)=0.130$ and $\a_s(M_Z)=0.118$, are used for LO and NLO predictions respectively. At NLO, we use $\a_s$ in two-loop formula as
\be
\dfrac{\a_s(\mu)}{4\pi} =\dfrac{1}{\b_0 \ln(\mu^2/\Lambda_{QCD}^2)} - \dfrac{\b_1 \ln \ln(\mu^2/\Lambda_{QCD}^2)}{\b_0^3 \ln^2(\mu^2/\Lambda_{QCD}^2)},
\ee
where $\b_1=34C_A^2/3-4(C_F+5C_A/3)T_F n_f$ is two loop coefficient of the QCD beta function.
For the heavy quark mass and the wave function at the
origin, $m_c=1.5 \gev$ and $|R_s(0)|^2 =0.810 \gev^3$ are used for $\jpsi$, and $m_b=4.75 \gev$ and $|R_s(0)|^2 =0.479 \gev^3$ are used for $\Upsilon$.
To choose the renormalization scale $\mu_r$ and the factorization scale $\mu_f$ in the calculations is an important issue and it causes the uncertainties for the calculation.
We choose $\mu=\mu_r=\mu_f=\sqrt{(2m_Q)^2+p_t^2}$ as the default choice in the calculation with $m_Q$ being $m_c$ and $m_b$ for $\jpsi$ and $\Upsilon$ respectively.
The center-of-mass energies are chosen as 1.98 TeV at Tevatron and 14 TeV at LHC.
The two phase space cutoffs $\d_s$ and $\d_c$ are chosen as $\d_s=10^{-3}$
and $\d_c=\d_s/50$ as default choice.
To check the independence of the final results on the two cutoffs, different values of $\d_s$ and $\d_c$ are used, where $\d_s$ can be as small as $\d_s=10^{-5}$. And the invariance is observed within the error tolerance of less than one percent.

It is known that the perturbative expansion cannot be applicable to the regions
with small transverse momentum and large rapidity of $\jpsi$ or $\Upsilon$. Therefore,
$P_t>3$ GeV are used for all the calculations. For rapidity cut at Tevatron,
we choose the same cut condition as the experiments at Tevatron
\cite{Abulencia:2007us,Abazov:2008za}: $|y|<0.6$ for $\jpsi$
and $|y|<1.8$ for $\Upsilon$. To follow the same cut condition used in
Ref.~\cite{Campbell:2007ws},
we choose $|y|<3$ for all calculation at LHC and another calculation of $\jpsi$ production
at Tevatron. All the cut conditions are explicitly expressed for each result.
\begin{figure}
\center{
\includegraphics*[scale=0.40]{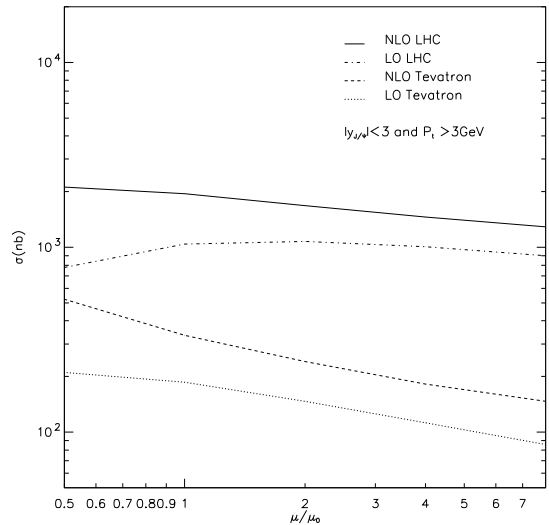}
\caption {\label{fig:jpsi_total}Total cross section of
$\jpsi$ production at Tevatron and LHC, as function of
the renormalization and factorization scale with
$\mu_r=\mu_f=\mu$ and $\mu_0=\sqrt{(2m_c)^2+p_t^2}$.}}
\end{figure}

The dependences of the total cross section at the renormalization scale $\mu_r$ and
factorization scale $\mu_f$ are presented in Fig.~\ref{fig:jpsi_total}. Since the contribution from the subprocess $gg\rightarrow \jpsi c\overline{c}$ is less than $10\%$ of the total result at NLO in the whole region of $\mu$, it gives almost
same plot as the Fig. 3 in ref.~\cite{Campbell:2007ws} which does not included the contribution.
The results show that the NLO QCD corrections boost the total cross section by a factor of about 2
at the default choice of the scales $\mu=\mu_0=\sqrt{(2m_c)^2+p_t^2}$. one can find that
the scale dependence at NLO is not improved for $\jpsi$.
\begin{figure}
\center{
\includegraphics*[scale=0.40]{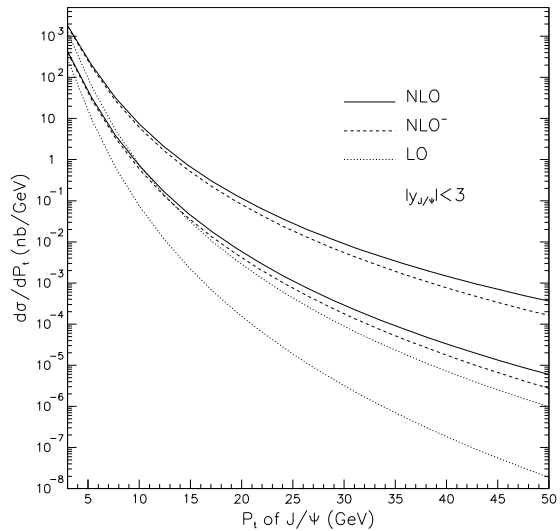}
\caption {\label{fig:jpsi_pt1}Transverse momentum distribution of $\jpsi$ production
at LHC (upper curves) and Tevatron (lower curves).
$\rm NLO^-$ denotes result excluding contribution from subprocess
$gg\rightarrow \jpsi c\overline{c}$.}}
\end{figure}
\begin{figure}
\center{
\includegraphics*[scale=0.40]{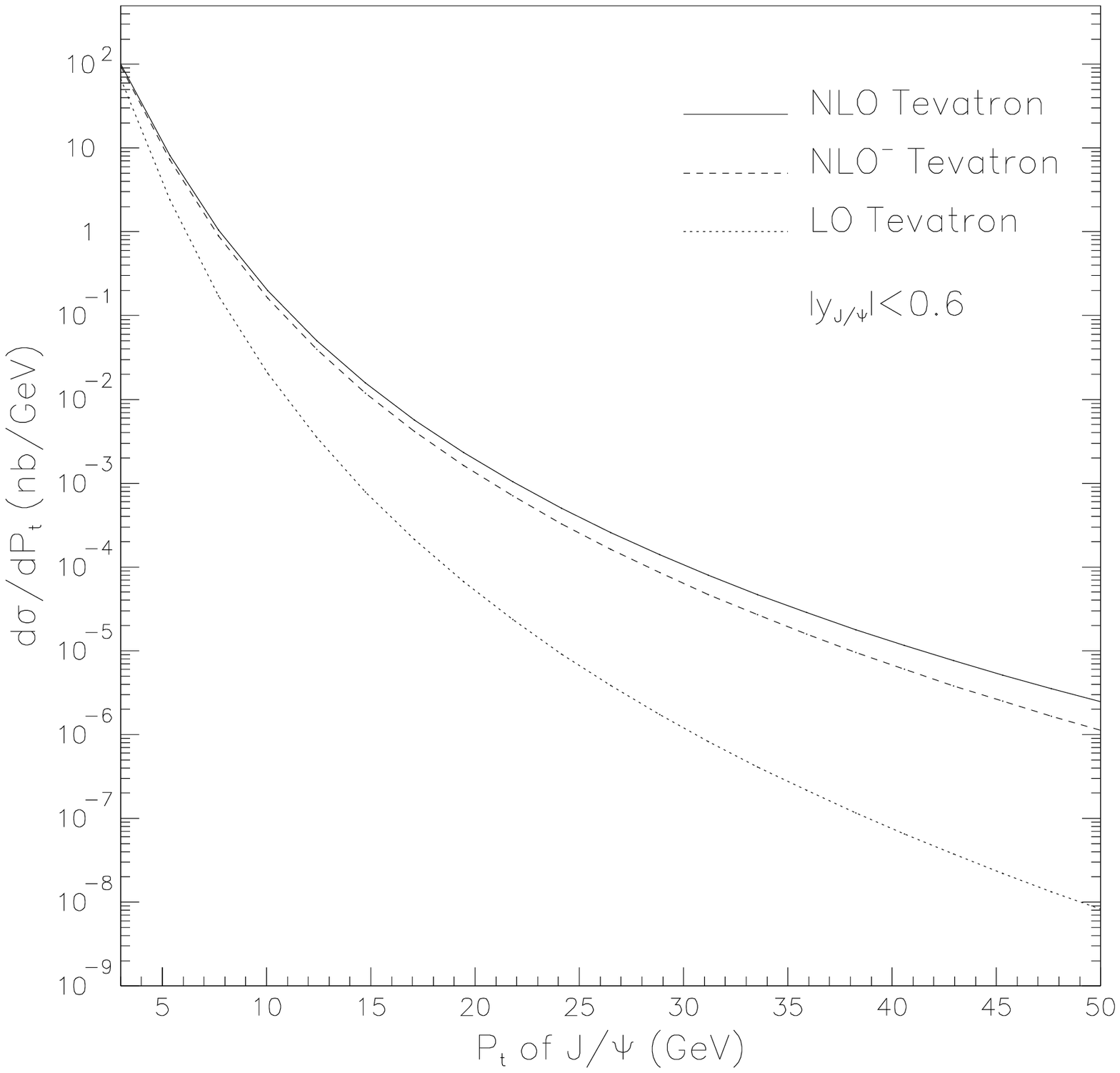}
\caption {\label{fig:jpsi_pt2}Transverse momentum distribution of $\jpsi$ production at Tevatron.  $\rm NLO^-$ denotes result excluding contribution from subprocess
$gg\rightarrow \jpsi c\overline{c}$.}}
\end{figure}
\begin{figure}
\center{
\includegraphics*[scale=0.40]{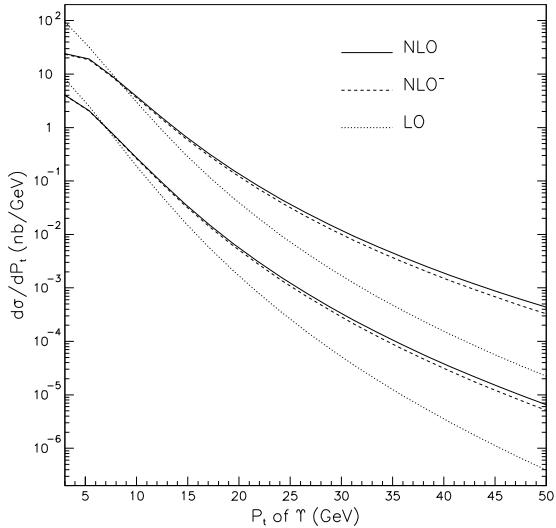}
\caption {\label{fig:upsilon_pt}Transverse momentum distribution of $\Upsilon$ production at LHC (upper curves)
and Tevatron (lower curves). $|y_\Upsilon|<1.8$ and $|y_\Upsilon|<3$ is taken for Tevatron and LHC respectively.
$\rm NLO^-$ denotes result excluding contribution from subprocess
$gg\rightarrow \Upsilon b\overline{b}$.}}
\end{figure}

In Figs.~\ref{fig:jpsi_pt1}, \ref{fig:jpsi_pt2} and
\ref{fig:upsilon_pt}, the $p_t$ distribution of $\jpsi$ and
$\Upsilon$ is shown.  It is easy to see that the contribution of NLO
correction becomes larger as $p_t$ increases, and in high $p_t$
region, the NLO prediction is 2-3 order of magnitude larger than the
LO one. As already known, the contribution from subprocesses $gg
\rightarrow \jpsi c\overline{c}$ or $gg\rightarrow \Upsilon
b\overline{b}$, which is also of $\co(\a_s)$, is large at high $p_t$
region. In order to compare with results in ref.~\cite{Campbell:2007ws} and also to see how large is the contribution, result excluding this contribution is shown in
the figures as $\rm NLO^-$. And we could see from the figures that
the contribution from $gg\rightarrow \Upsilon b\overline{b}$ in
$\Upsilon$ production is less than that from $gg \rightarrow \jpsi
c\overline{c}$ in $\jpsi$ case.

\begin{figure*}
\center{
\includegraphics*[scale=0.35]{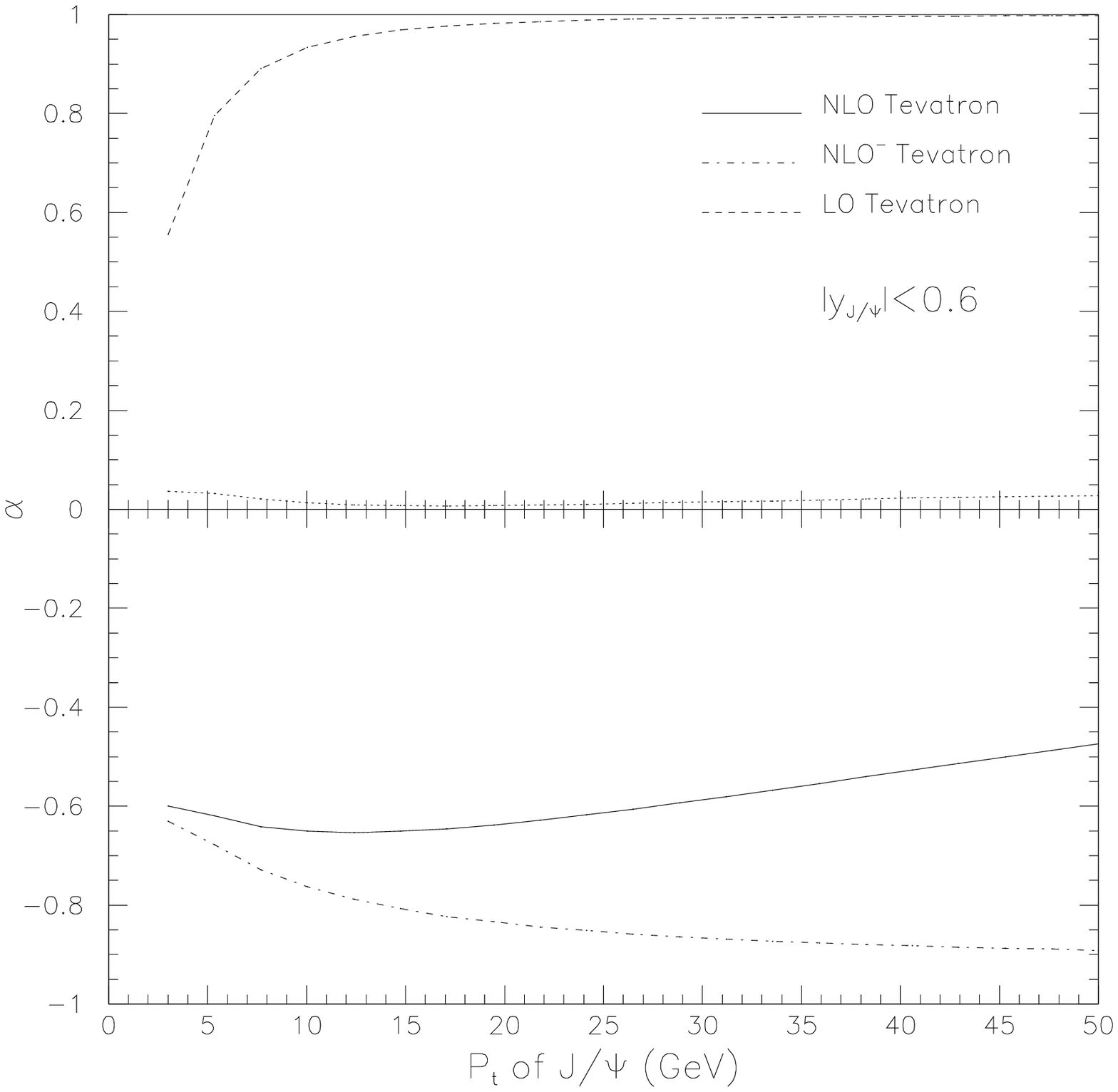}
~~~~~~~~~~~~~~~~~~
\includegraphics*[scale=0.35]{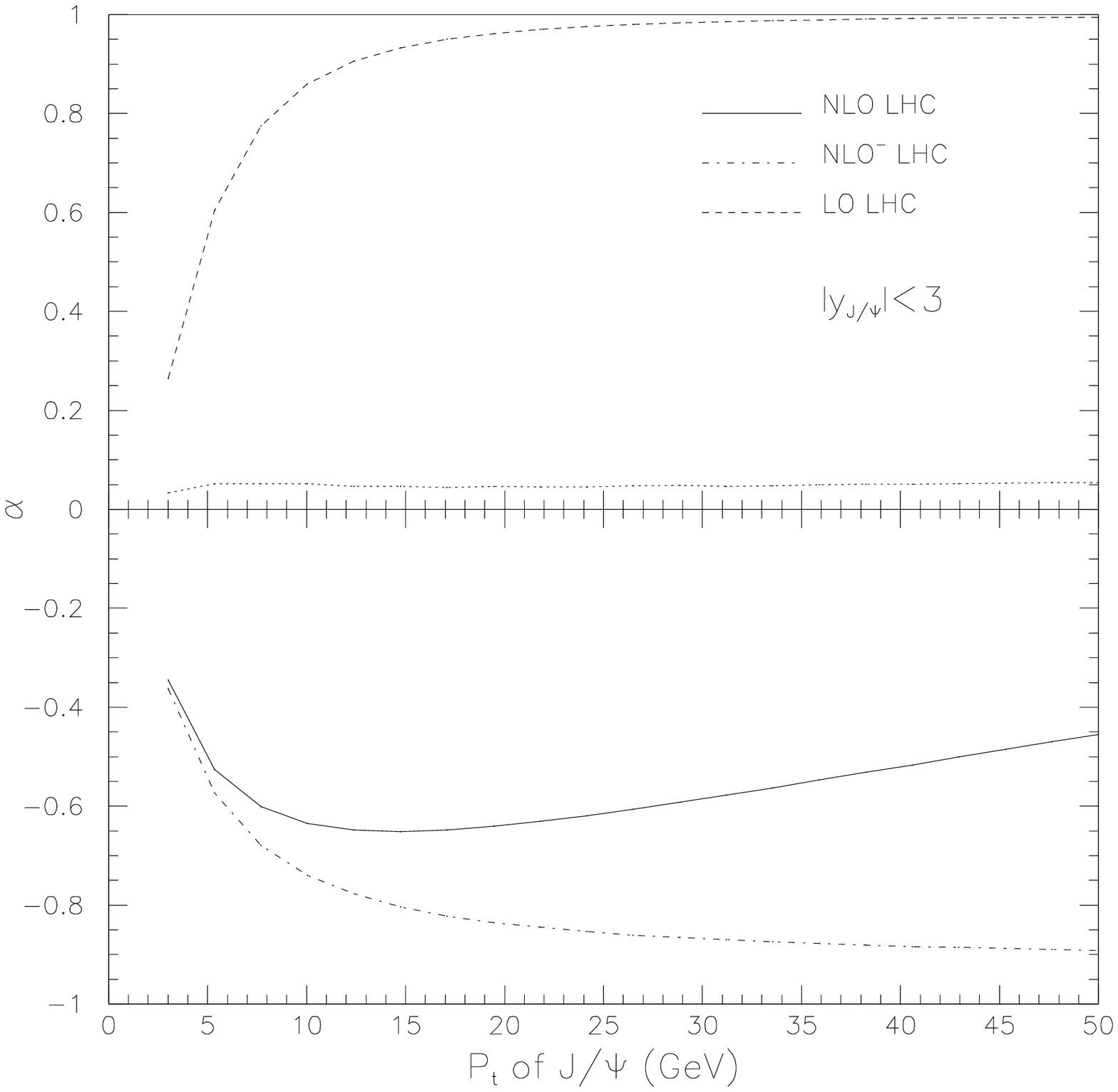}
\caption {\label{fig:jpsi_polar}Transverse momentum distribution of $\jpsi$ polarization
at Tevatron and LHC.
$\rm NLO^-$ denotes result excluding contribution from subprocess
$gg\rightarrow \jpsi c\overline{c}$
and the process itself is represented by the unlabeled dotted line.}}
\end{figure*}
\begin{figure*}
\center{
\includegraphics*[scale=0.35]{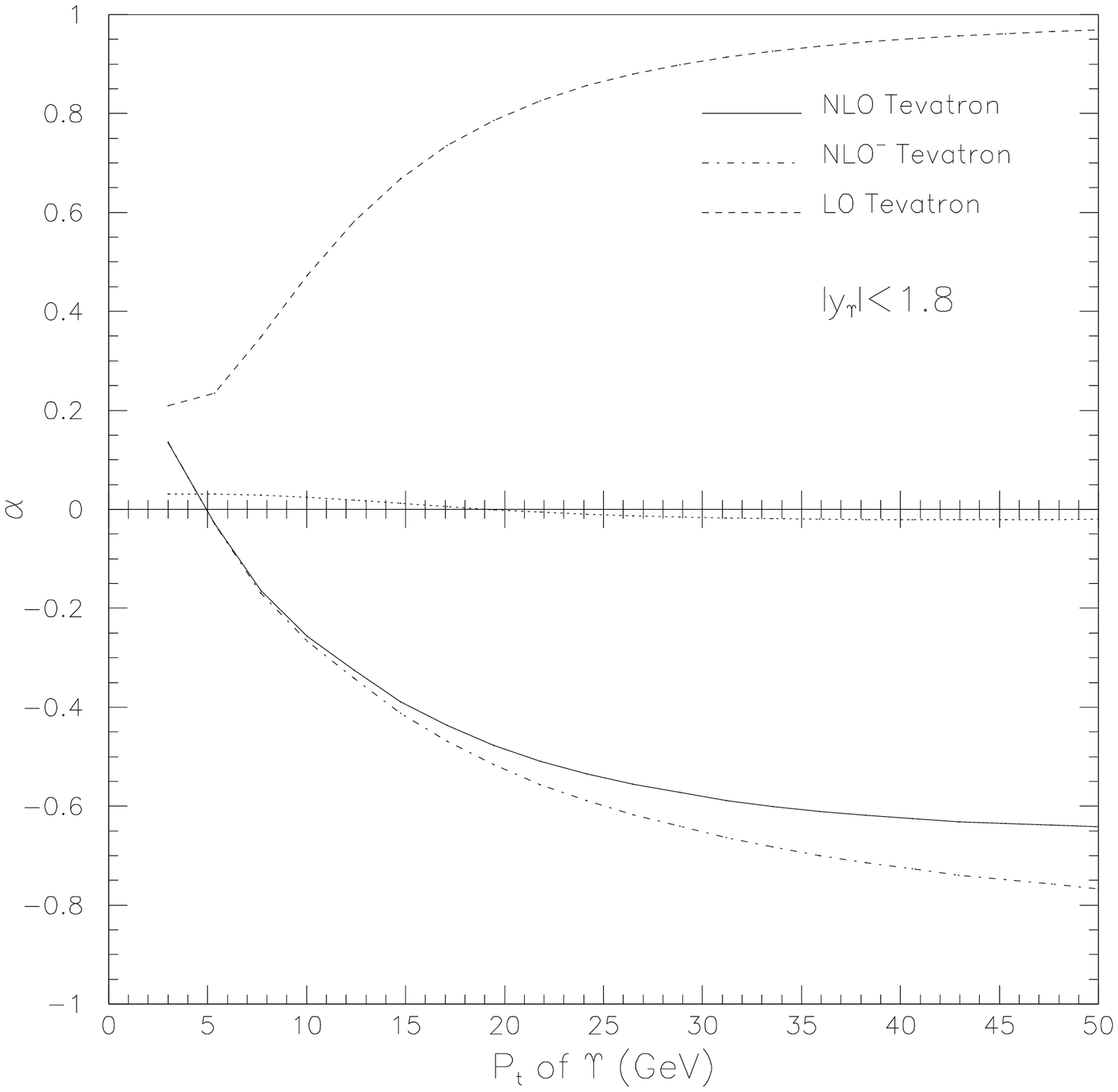}
~~~~~~~~~~~~~~~~~~
\includegraphics*[scale=0.35]{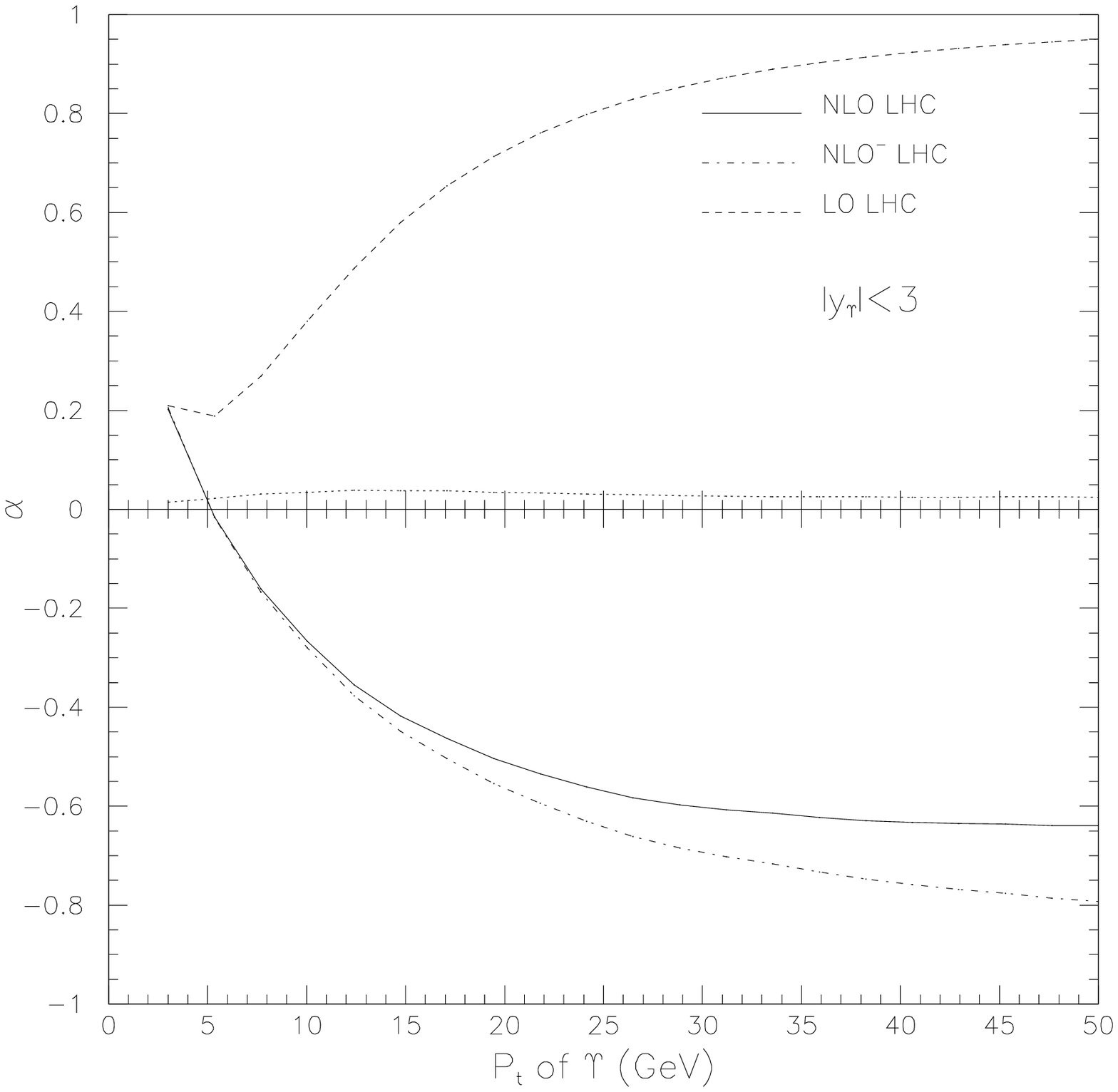}
\caption {\label{fig:upsilon_polar}Transverse momentum distribution of $\Upsilon$ polarization at Tevatron and LHC.
$\rm NLO^-$ denotes result excluding contribution from subprocess
$gg\rightarrow \Upsilon b\overline{b}$
and the process itself is represented by the unlabeled dotted line.}}
\end{figure*}
The $p_t$ distribution of $\jpsi$ and $\Upsilon$ polarization factor
$\a$ is presented in Fig.~\ref{fig:jpsi_polar} and
Fig.~\ref{fig:upsilon_polar}. We can see in the figures that $\a$ is always positive and
becomes closer to 1 as $p_t$ increases at LO, and this figure means that
the transverse polarization is more than the longitudinal one and
even becomes dominant in high $p_t$ region. But there is dramatical
change when the NLO QCD corrections are taken into account. For
$\jpsi$,  $\a$ is always negative and becomes closer to -0.5 as $p_t$
increases, this new figure indicates that the longitudinal
polarization is always more than the transverse one and even becomes
dominant in high $p_t$ region. Meanwhile the $\jpsi$ polarization in
subprocess $gg\rightarrow \jpsi c\overline{c}$  is near zero. By
including contribution of this subprocess, the total result shown in the
left diagram of Fig.~\ref{fig:jpsi_polar} is closer
to the experimental result. For $\Upsilon$,  $\a$ varies from
positive to negative and becomes closer to -0.6 as $p_t$ increases,
this indicates that the longitudinal polarization becomes more and
more, and even becomes dominant in high $p_t$ region. The $\Upsilon$
polarization of subprocess $gg\rightarrow \Upsilon b\overline{b}$ is
also near zero. But from the total result in Fig.~\ref{fig:upsilon_polar} we can see that this subprocess contributes less than the corresponding one in case of $\jpsi$. Also, we
find that the contribution from light quarks affects the $p_t$
distribution of polarization less than $10\%$. When compare the
figures for $\jpsi$ with those for $\Upsilon$, we can see that they
are very similar to each other except that $\alpha$ is higher and
even becomes positive in lower $p_t$ region for $\Upsilon$. It could
be understand by extending the curves for $\jpsi$ to lower $p_t$,
because for a certain $p_t$ value in $\Upsilon$ production, it
corresponds a lower $p_t$ in $\jpsi$ production by just considering
the energy scale.

By comparing the experimental measurements for $\jpsi$
\cite{Abulencia:2007us,Abulencia:2005us} and for $\Upsilon$
\cite{Acosta:2001gv,Abazov:2008za} at Tevatron with above results,
we could see that, although NLO corrections can boost the transverse
momentum distribution of $\jpsi$ and $\Upsilon$ very much, it is still 
an order of mangitude smaller than the experimental data. The color octet channels
are still needed to explain the $p_t$ distribution. Thus the NLO
prediction for the polarization of direct $\jpsi$ and $\Upsilon$ via
color singlet channel could not be used to compare with experimental
data.

We can write contribution of each channel as \be
\sigma^i=C_\epsilon\left(C_2^i\dfrac{1}{\epsilon^2} +
C_1^i\dfrac{1}{\epsilon} +C_0^i \right) \ee where the overall $\e$
dependent factor \be
C_\epsilon=\left[\dfrac{1}{(1-\epsilon)^2}\dfrac{4\pi\mu_r^2}{(2m_c)^2}\right]^\epsilon
e^{-\epsilon\gamma_E}, \ee and the term $1/{(1-\epsilon)^2}$ is from
the gluon spin average factor $1/{(n-2)}$. When all the
contributions are summed up, we have $\sum C_2^i=0$ and $\sum
C_1^i=0$. Thus $C_\epsilon$ comes back to 1 and we have our result
as $\sum C_0^i$. In Table.~\ref{channel}, $C_0^i$ is given. It
should be careful that the $A^{sc}_0(g\rightarrow gg)$ term has been
put into the $gg\rightarrow\jpsi+gg$ channel even if it contains a
term proportional to the number of active flavors $n_f$.

\begin{table}[htbp]
\begin{center}
\begin{tabular}{|c|c|c|c|c|}
\hline\hline i & process & $C_{0}^{i}$($10^{2}$nb)&$C_0^i/\sigma^B$ &fraction\\
\hline 1 & $gg\rightarrow {J/\psi }g$ & 0.4061$\pm$0.0006&0.2174 &0.1056\\
\hline 2 & $gg\rightarrow {J/\psi }gg$ & 2.47$\pm$0.04&1.32 &0.64\\
\hline 3 & $gg\rightarrow {J/\psi }q\bar{q}$ & 0.133$\pm$0.001 &0.071&0.035\\
\hline 4 & $gq\rightarrow {J/\psi }gq$ & 0.582$\pm$0.001 &0.312&0.152\\
\hline 5 & $gg\rightarrow {J/\psi }c\bar{c}$ & 0.2583$\pm$0.0003&0.1382 &0.0672\\
\hline $\sum$ & $p\bar{p}\rightarrow \jpsi +X $&3.84$\pm$0.04&2.06&1.00
\\  \hline\hline
\end{tabular}
\caption{lists of contributions from each channel to the NLO total cross section of
$\jpsi$ hadronproduction at Tevatron in the region $p_t>3$ GeV and
$|y_{\jpsi}|<3$. We have set $\mu_r=\mu_f=\mu_0$. Corresponding
result for $\sigma^B$ is $ 1.8682\times 10^2$ nb.} \label{channel}
\end{center}
\end{table}
\section{Conclusion and Discussion}
We have calculated the NLO QCD corrections to the $\jpsi$ and $\Upsilon$ hadronproduction at Tevatron  and LHC. Dimensional regularization is applied to deal with the UV and IR singularities in the calculation, and the Coulomb singularity is isolated by a small relative velocity $v$ between the quark pair in the
meson and absorbed into the bound state wave function. To deal
with the soft and collinear singularities in the real corrections, the two-cutoff phase space slicing method is used.
By summing over all the contributions, a result which is UV, IR and Coulomb finite is obtained.

Numerically, we obtain a K factor of total cross section
(ratio of NLO to LO) of about 2 for $\jpsi$. The transverse momentum distributions
of $\jpsi$ and $\Upsilon$ are presented and they show that the NLO corrections
increase the differential cross sections more as $p_t$ becomes larger
and eventually can enhance it by 2 or 3 orders in magnitude at $p_t=50 \gev$.
It confirms the calculation by Campbell, Maltoni and Tramontano \cite{Campbell:2007ws}. The real correction subprocesses $gg\rightarrow \jpsi c\overline{c}$ and  $gg\rightarrow \Upsilon b\overline{b}$ are also calculated and the results are in agreement with those of Ref.~\cite{Qiao:2003ba,Artoisenet:2007xi}.

The NLO contributions to $\jpsi$ polarization is studied and our
results indicate that the $\jpsi$ polarization is dramatically
changed from more transverse polarization at LO into more
longitudinal polarization at NLO. All the results can be directly
applied to $\psi^\prime$ production by multiplying a factor
$\mopsip/\mopsi$. The NLO contributions to $\Upsilon$ polarization
is also studied and presented for the first time. Our results
indicates that at NLO, the polarization of $\Upsilon$ decreases
gradually from near 0.2 to -0.6 as $p_t$ increases from 3 GeV to 50
GeV. Namely, the $p_t$ distribution of the polarization status
behaves almost the same as that for $\jpsi$ except that the NLO
result is also transverse polarization at small $p_t$ range. Since
the fact that contribution via color-octet states is much less in
$\Upsilon$ production than that in $\jpsi$ case, our new result for
$\Upsilon$ polarization plays an important role in understanding the
experimental data. And even though our calculation results in a more
longitudinal polarization state than the recent experimental result
for $\jpsi$ \cite{Abulencia:2007us} and $\Upsilon$
\cite{Abazov:2008za} at Tevatron, it raises a hope to solve the
large discrepancy between LO theoretical predication and
experimental measurement on $\jpsi$ and  $\Upsilon$ polarization,
and suggests that the next important step is to calculate the NLO
corrections to hadronproduction of color octet state $\jpsi^{(8)}$
and $\Upsilon^{(8)}$.  By re-fixing the color-octet matrix elements,
we will see what an involvement of the NLO QCD corrections can
induce for the polarization of $\jpsi$ and $\Upsilon$.

\begin{acknowledgments}
This work was supported by the National Natural Science Foundation of
China (No.~10475083) and by the Chinese Academy of Sciences under
Project No. KJCX3-SYW-N2.
\end{acknowledgments}
\appendix
\bibliography{twain}

\begin{thebibliography}{33}
\expandafter\ifx\csname natexlab\endcsname\relax\def\natexlab#1{#1}\fi
\expandafter\ifx\csname bibnamefont\endcsname\relax
  \def\bibnamefont#1{#1}\fi
\expandafter\ifx\csname bibfnamefont\endcsname\relax
  \def\bibfnamefont#1{#1}\fi
\expandafter\ifx\csname citenamefont\endcsname\relax
  \def\citenamefont#1{#1}\fi
\expandafter\ifx\csname url\endcsname\relax
  \def\url#1{\texttt{#1}}\fi
\expandafter\ifx\csname urlprefix\endcsname\relax\def\urlprefix{URL }\fi
\providecommand{\bibinfo}[2]{#2}
\providecommand{\eprint}[2][]{\url{#2}}

\bibitem{j.h.kuhn:79}
J.H. K\"uhn, J. Kaplan, and E.G.O. Safiani,
\npb{\bf 157}, 125 (1979);\\
C.H. Chang, \npb{\bf 172}, 425 (1980);\\
B. Guberina, J.H. K\"uhn, R.D. Peccei, and R. R\"uckl,
\npb{\bf 174}, 317 (1980);\\
E.L. Berger and D. Jones, \prd{\bf 23}, 1521 (1981);\\
R. Baier and R. R\"uckl, Z. Phys. C{\bf 19}, 251 (1983).

\bibitem{cdf1}
CDF Collaboration, F.~Abe {\it et al.}
Phys.\ Rev.\ Lett.\  {\bf 69}, 3704 (1992);
Phys.\ Rev.\ Lett.\  {\bf 79}, 572 (1997);
Phys.\ Rev.\ Lett.\  {\bf 79}, 578 (1997);

\bibitem[{\citenamefont{Braaten and Fleming}(1995)}]{Braaten:1994vv}
\bibinfo{author}{\bibfnamefont{E.}~\bibnamefont{Braaten}} \bibnamefont{and}
  \bibinfo{author}{\bibfnamefont{S.}~\bibnamefont{Fleming}},
  \bibinfo{journal}{Phys. Rev. Lett.} \textbf{\bibinfo{volume}{74}},
  \bibinfo{pages}{3327} (\bibinfo{year}{1995}).

\bibitem[{\citenamefont{Bodwin et~al.}(1995)\citenamefont{Bodwin, Braaten, and
  Lepage}}]{Bodwin:1994jh}
\bibinfo{author}{\bibfnamefont{G.~T.} \bibnamefont{Bodwin}},
  \bibinfo{author}{\bibfnamefont{E.}~\bibnamefont{Braaten}}, \bibnamefont{and}
  \bibinfo{author}{\bibfnamefont{G.~P.} \bibnamefont{Lepage}},
  \bibinfo{journal}{Phys. Rev.} \textbf{\bibinfo{volume}{D51}},
  \bibinfo{pages}{1125} (\bibinfo{year}{1995}).

\bibitem[{\citenamefont{Kniehl and Kramer}(1997)}]{Kniehl:1997fv}
\bibinfo{author}{\bibfnamefont{B.~A.} \bibnamefont{Kniehl}} \bibnamefont{and}
  \bibinfo{author}{\bibfnamefont{G.}~\bibnamefont{Kramer}},
  \bibinfo{journal}{Phys. Lett.} \textbf{\bibinfo{volume}{B413}},
  \bibinfo{pages}{416} (\bibinfo{year}{1997}).

\bibitem[{\citenamefont{Ko et~al.}(1996)\citenamefont{Ko, Lee, and
  Song}}]{Ko:1996xw}
\bibinfo{author}{\bibfnamefont{P.}~\bibnamefont{Ko}},
  \bibinfo{author}{\bibfnamefont{J.}~\bibnamefont{Lee}}, \bibnamefont{and}
  \bibinfo{author}{\bibfnamefont{H.~S.} \bibnamefont{Song}},
  \bibinfo{journal}{Phys. Rev.} \textbf{\bibinfo{volume}{D54}},
  \bibinfo{pages}{4312} (\bibinfo{year}{1996}).

\bibitem[{\citenamefont{Kramer}(1996)}]{kramer:1995nb}
\bibinfo{author}{\bibfnamefont{M.}~\bibnamefont{Kramer}},
  \bibinfo{journal}{Nucl. Phys.} \textbf{\bibinfo{volume}{B459}},
  \bibinfo{pages}{3} (\bibinfo{year}{1996}).

\bibitem[{\citenamefont{Kramer et~al.}(1995)\citenamefont{Kramer, Zunft,
  Steegborn, and Zerwas}}]{Kramer:1994zi}
\bibinfo{author}{\bibfnamefont{M.}~\bibnamefont{Kramer}},
  \bibinfo{author}{\bibfnamefont{J.}~\bibnamefont{Zunft}},
  \bibinfo{author}{\bibfnamefont{J.}~\bibnamefont{Steegborn}},
  \bibnamefont{and} \bibinfo{author}{\bibfnamefont{P.~M.}
  \bibnamefont{Zerwas}}, \bibinfo{journal}{Phys. Lett.}
  \textbf{\bibinfo{volume}{B348}}, \bibinfo{pages}{657} (\bibinfo{year}{1995}).

\bibitem[{\citenamefont{Amundson et~al.}(1997)\citenamefont{Amundson, Fleming,
  and Maksymyk}}]{Amundson:1996ik}
\bibinfo{author}{\bibfnamefont{J.}~\bibnamefont{Amundson}},
  \bibinfo{author}{\bibfnamefont{S.}~\bibnamefont{Fleming}}, \bibnamefont{and}
  \bibinfo{author}{\bibfnamefont{I.}~\bibnamefont{Maksymyk}},
  \bibinfo{journal}{Phys. Rev.} \textbf{\bibinfo{volume}{D56}},
  \bibinfo{pages}{5844} (\bibinfo{year}{1997}).

\bibitem[{\citenamefont{Cacciari and Kramer}(1996)}]{cacciari:1996dg}
\bibinfo{author}{\bibfnamefont{M.}~\bibnamefont{Cacciari}} \bibnamefont{and}
  \bibinfo{author}{\bibfnamefont{M.}~\bibnamefont{Kramer}},
  \bibinfo{journal}{Phys. Rev. Lett.} \textbf{\bibinfo{volume}{76}},
  \bibinfo{pages}{4128} (\bibinfo{year}{1996}).

\bibitem{kramer:2001} 
M. Kramer, Prog. Part. Nucl. Phys. {\bf 47}, 141 (2001); 
J. P. Lansberg, Int. J. Mod. Phys. A {bf 21}, 3857 (2006).

\bibitem{hera:h1}
H1 Collaboration, Contributed Paper 157aj, International Europhysics
Conference on High Energy Physics (EPS99), Tampere, Finland, 1999.

\bibitem{hera:zeus}
ZEUS Collaboration, Contributed Paper 851, International
Conference on High Energy Physics (ICHEP2000), Osaka, Japan, 2000.


\bibitem[{\citenamefont{Klasen et~al.}(2002)\citenamefont{Klasen, Kniehl,
  Mihaila, and Steinhauser}}]{Klasen:2001cu}
\bibinfo{author}{\bibfnamefont{M.}~\bibnamefont{Klasen}},
  \bibinfo{author}{\bibfnamefont{B.~A.} \bibnamefont{Kniehl}},
  \bibinfo{author}{\bibfnamefont{L.~N.} \bibnamefont{Mihaila}},
  \bibnamefont{and}
  \bibinfo{author}{\bibfnamefont{M.}~\bibnamefont{Steinhauser}},
  \bibinfo{journal}{Phys. Rev. Lett.} \textbf{\bibinfo{volume}{89}},
  \bibinfo{pages}{032001} (\bibinfo{year}{2002}).

\bibitem[{\citenamefont{de~Boer and Sander}(2004)}]{deBoer:2003xm}
\bibinfo{author}{\bibfnamefont{W.}~\bibnamefont{de~Boer}} \bibnamefont{and}
  \bibinfo{author}{\bibfnamefont{C.}~\bibnamefont{Sander}},
  \bibinfo{journal}{Phys. Lett.} \textbf{\bibinfo{volume}{B585}},
  \bibinfo{pages}{276} (\bibinfo{year}{2004}).

\bibitem[{\citenamefont{Qiao and Wang}(2004)}]{Qiao:2003ba}
\bibinfo{author}{\bibfnamefont{C.-F.} \bibnamefont{Qiao}} \bibnamefont{and}
  \bibinfo{author}{\bibfnamefont{J.-X.} \bibnamefont{Wang}},
  \bibinfo{journal}{Phys. Rev.} \textbf{\bibinfo{volume}{D69}},
  \bibinfo{pages}{014015} (\bibinfo{year}{2004}).

\bibitem[{\citenamefont{Hagiwara et~al.}(2007)\citenamefont{Hagiwara, Qi, Qiao,
  and Wang}}]{Hagiwara:2007bq}
\bibinfo{author}{\bibfnamefont{K.}~\bibnamefont{Hagiwara}},
  \bibinfo{author}{\bibfnamefont{W.}~\bibnamefont{Qi}},
  \bibinfo{author}{\bibfnamefont{C.~F.} \bibnamefont{Qiao}}, \bibnamefont{and}
  \bibinfo{author}{\bibfnamefont{J.~X.} \bibnamefont{Wang}}
  (\bibinfo{year}{2007}), \eprint{arXiv:0705.0803 [hep-ph]}.

\bibitem[{\citenamefont{Braaten and Lee}(2003)}]{Braaten:2002fi}
\bibinfo{author}{\bibfnamefont{E.}~\bibnamefont{Braaten}} \bibnamefont{and}
  \bibinfo{author}{\bibfnamefont{J.}~\bibnamefont{Lee}},
  \bibinfo{journal}{Phys. Rev.} \textbf{\bibinfo{volume}{D67}},
  \bibinfo{pages}{054007} (\bibinfo{year}{2003}).

\bibitem[{\citenamefont{Liu et~al.}(2003)\citenamefont{Liu, He, and
  Chao}}]{Liu:2002wq}
\bibinfo{author}{\bibfnamefont{K.-Y.} \bibnamefont{Liu}},
  \bibinfo{author}{\bibfnamefont{Z.-G.} \bibnamefont{He}}, \bibnamefont{and}
  \bibinfo{author}{\bibfnamefont{K.-T.} \bibnamefont{Chao}},
  \bibinfo{journal}{Phys. Lett.} \textbf{\bibinfo{volume}{B557}},
  \bibinfo{pages}{45} (\bibinfo{year}{2003}).

\bibitem[{\citenamefont{Hagiwara et~al.}(2003)\citenamefont{Hagiwara, Kou, and
  Qiao}}]{Hagiwara:2003cw}
\bibinfo{author}{\bibfnamefont{K.}~\bibnamefont{Hagiwara}},
  \bibinfo{author}{\bibfnamefont{E.}~\bibnamefont{Kou}}, \bibnamefont{and}
  \bibinfo{author}{\bibfnamefont{C.-F.} \bibnamefont{Qiao}},
  \bibinfo{journal}{Phys. Lett.} \textbf{\bibinfo{volume}{B570}},
  \bibinfo{pages}{39} (\bibinfo{year}{2003}).

\bibitem[{\citenamefont{Abe et~al.}(2002)}]{Abe:2002rb}
\bibinfo{author}{\bibfnamefont{K.}~\bibnamefont{Abe}} \bibnamefont{et~al.}
  (\bibinfo{collaboration}{Belle}), \bibinfo{journal}{Phys. Rev. Lett.}
  \textbf{\bibinfo{volume}{89}}, \bibinfo{pages}{142001}
  (\bibinfo{year}{2002}).

\bibitem[{\citenamefont{Aubert et~al.}(2005)}]{Aubert:2005tj}
\bibinfo{author}{\bibfnamefont{B.}~\bibnamefont{Aubert}} \bibnamefont{et~al.}
  (\bibinfo{collaboration}{BABAR}), \bibinfo{journal}{Phys. Rev.}
  \textbf{\bibinfo{volume}{D72}}, \bibinfo{pages}{031101}
  (\bibinfo{year}{2005}).

\bibitem[{\citenamefont{Zhang et~al.}(2006)\citenamefont{Zhang, Gao, and
  Chao}}]{Zhang:2005ch}
\bibinfo{author}{\bibfnamefont{Y.-J.} \bibnamefont{Zhang}},
  \bibinfo{author}{\bibfnamefont{Y.-j.} \bibnamefont{Gao}}, \bibnamefont{and}
  \bibinfo{author}{\bibfnamefont{K.-T.} \bibnamefont{Chao}},
  \bibinfo{journal}{Phys. Rev. Lett.} \textbf{\bibinfo{volume}{96}},
  \bibinfo{pages}{092001} (\bibinfo{year}{2006}).

\bibitem[{\citenamefont{Gong and Wang}(2007)}]{jxwang:2007je}
\bibinfo{author}{\bibfnamefont{B.}~\bibnamefont{Gong}} \bibnamefont{and}
  \bibinfo{author}{\bibfnamefont{J.-X.} \bibnamefont{Wang}},
  \bibinfo{journal}{Phys. Rev.} \textbf{\bibinfo{volume}{D77}},
  \bibinfo{pages}{054028} (\bibinfo{year}{2008}), \eprint{arXiv:0712.4220 [hep-ph]}.

\bibitem[{\citenamefont{Gong and Wang}(2008)}]{Gong:2008ce}
\bibinfo{author}{\bibfnamefont{B.}~\bibnamefont{Gong}} \bibnamefont{and}
  \bibinfo{author}{\bibfnamefont{J.-X.} \bibnamefont{Wang}},
  \bibinfo{journal}{Phys. Rev. Lett.} \textbf{\bibinfo{volume}{100}},
  \bibinfo{pages}{181803} (\bibinfo{year}{2008}), \eprint{arXiv:0801.0648 [hep-ph]}.

\bibitem[{\citenamefont{He et~al.}(2007)\citenamefont{He, Fan, and
  Chao}}]{He:2007te}
\bibinfo{author}{\bibfnamefont{Z.-G.} \bibnamefont{He}},
  \bibinfo{author}{\bibfnamefont{Y.}~\bibnamefont{Fan}}, \bibnamefont{and}
  \bibinfo{author}{\bibfnamefont{K.-T.} \bibnamefont{Chao}},
  \bibinfo{journal}{Phys. Rev.} \textbf{\bibinfo{volume}{D75}},
  \bibinfo{pages}{074011} (\bibinfo{year}{2007}).

\bibitem[{\citenamefont{Zhang and Chao}(2007)}]{Zhang:2006ay}
\bibinfo{author}{\bibfnamefont{Y.-J.} \bibnamefont{Zhang}} \bibnamefont{and}
  \bibinfo{author}{\bibfnamefont{K.-T.} \bibnamefont{Chao}},
  \bibinfo{journal}{Phys. Rev. Lett.} \textbf{\bibinfo{volume}{98}},
  \bibinfo{pages}{092003} (\bibinfo{year}{2007}).
\bibitem[{\citenamefont{Zhang et~al.}(2008)\citenamefont{Zhang, Ma, and
  Chao}}]{Zhang:2008gp}
\bibinfo{author}{\bibfnamefont{Y.-J.} \bibnamefont{Zhang}},
  \bibinfo{author}{\bibfnamefont{Y.-Q.} \bibnamefont{Ma}}, \bibnamefont{and}
  \bibinfo{author}{\bibfnamefont{K.-T.} \bibnamefont{Chao}}
  (\bibinfo{year}{2008}), \eprint{arXiv:0802.3655 [hep-ph]}.

\bibitem{beneke:96yr}
M. Beneke and I.Z. Rothstein, \plb{\bf 372}, 157 (1996),
[Erratum-ibid. B{\bf 389}, 769 (1996)];
M. Beneke and M. Kr\"amer, \prd{\bf 55}, 5269 (1997).

\bibitem{braaten:99yr} 
E. Braaten, B.A. Kniehl, and J. Lee, \prd{\bf 62}, 094005 (2000);
B. A. Kniehl and J. Lee, \prd{\bf 62}, 114027 (2000);

\bibitem[{\citenamefont{Leibovich}(1997)}]{Leibovich:1996pa}
\bibinfo{author}{\bibfnamefont{A.~K.} \bibnamefont{Leibovich}},
  \bibinfo{journal}{Phys. Rev.} \textbf{\bibinfo{volume}{D56}},
  \bibinfo{pages}{4412} (\bibinfo{year}{1997}).

\bibitem[{\citenamefont{Abulencia et~al.}(2007)}]{Abulencia:2007us}
\bibinfo{author}{\bibfnamefont{A.}~\bibnamefont{Abulencia}}
  \bibnamefont{et~al.} (\bibinfo{collaboration}{CDF}), \bibinfo{journal}{Phys.
  Rev. Lett.} \textbf{\bibinfo{volume}{99}}, \bibinfo{pages}{132001}
  (\bibinfo{year}{2007}).

\bibitem{Abazov:2008za}
  V.~M.~Abazov {\it et al.}  [D0 Collaboration],
  arXiv:0804.2799 [hep-ex].

\bibitem{Braaten:2000gw}
  E.~Braaten and J.~Lee,
  Phys.\ Rev.\  D {\bf 63}, 071501 (2001)

\bibitem{Haberzettl:2007kj}
H. Haberzettl and J. P. Lansberg,
 Phys.\ Rev.\ Lett.\  {\bf 100}, 032006 (2008)

\bibitem[{\citenamefont{Campbell et~al.}(2007)\citenamefont{Campbell, Maltoni,
  and Tramontano}}]{Campbell:2007ws}
\bibinfo{author}{\bibfnamefont{J.}~\bibnamefont{Campbell}},
  \bibinfo{author}{\bibfnamefont{F.}~\bibnamefont{Maltoni}}, \bibnamefont{and}
  \bibinfo{author}{\bibfnamefont{F.}~\bibnamefont{Tramontano}},
  \bibinfo{journal}{Phys. Rev. Lett.} \textbf{\bibinfo{volume}{98}},
  \bibinfo{pages}{252002} (\bibinfo{year}{2007}).

\bibitem[{\citenamefont{Artoisenet et~al.}(2007)\citenamefont{Artoisenet,
  Lansberg, and Maltoni}}]{Artoisenet:2007xi}
\bibinfo{author}{\bibfnamefont{P.}~\bibnamefont{Artoisenet}},
  \bibinfo{author}{\bibfnamefont{J.~P.} \bibnamefont{Lansberg}},
  \bibnamefont{and} \bibinfo{author}{\bibfnamefont{F.}~\bibnamefont{Maltoni}},
  \bibinfo{journal}{Phys. Lett.} \textbf{\bibinfo{volume}{B653}},
  \bibinfo{pages}{60} (\bibinfo{year}{2007}).

\bibitem{Gong:2008sn}
\bibinfo{author}{\bibfnamefont{B.}~\bibnamefont{Gong}} \bibnamefont{and}
  \bibinfo{author}{\bibfnamefont{J.-X.} \bibnamefont{Wang}},
  \bibinfo{journal}{Phys. Rev. Lett.} \textbf{\bibinfo{volume}{100}},
  \bibinfo{pages}{232001} (\bibinfo{year}{2008}), \eprint{arXiv:0802.3727 [hep-ph]}.

\bibitem[{\citenamefont{Wang}(2004)}]{FDC}
\bibinfo{author}{\bibfnamefont{J.-X.} \bibnamefont{Wang}},
  \bibinfo{journal}{Nucl. Instrum. Meth.} \textbf{\bibinfo{volume}{A534}},
  \bibinfo{pages}{241} (\bibinfo{year}{2004}).

\bibitem[{\citenamefont{Klasen et~al.}(2005)\citenamefont{Klasen, Kniehl,
  Mihaila, and Steinhauser}}]{Klasen:2004tz}
\bibinfo{author}{\bibfnamefont{M.}~\bibnamefont{Klasen}},
  \bibinfo{author}{\bibfnamefont{B.~A.} \bibnamefont{Kniehl}},
  \bibinfo{author}{\bibfnamefont{L.~N.} \bibnamefont{Mihaila}},
  \bibnamefont{and}
  \bibinfo{author}{\bibfnamefont{M.}~\bibnamefont{Steinhauser}},
  \bibinfo{journal}{Nucl. Phys.} \textbf{\bibinfo{volume}{B713}},
  \bibinfo{pages}{487} (\bibinfo{year}{2005}).
  
\bibitem{Passarino:1978jh}
\bibinfo{author}{\bibfnamefont{G.}~\bibnamefont{Passarino}} \bibnamefont{and}
  \bibinfo{author}{\bibfnamefont{M.~J.~G.} \bibnamefont{Veltman}},
  \bibinfo{journal}{Nucl. Phys.} \textbf{\bibinfo{volume}{B160}},
  \bibinfo{pages}{151} (\bibinfo{year}{1979}).

\bibitem[{\citenamefont{Harris and Owens}(2002)}]{Harris:2001sx}
\bibinfo{author}{\bibfnamefont{B.~W.} \bibnamefont{Harris}} \bibnamefont{and}
  \bibinfo{author}{\bibfnamefont{J.~F.} \bibnamefont{Owens}},
  \bibinfo{journal}{Phys. Rev.} \textbf{\bibinfo{volume}{D65}},
  \bibinfo{pages}{094032} (\bibinfo{year}{2002}).

\bibitem[{\citenamefont{Collins et~al.}(1985)\citenamefont{Collins, Soper, and
  Sterman}}]{Collins:1985ue}
\bibinfo{author}{\bibfnamefont{J.~C.} \bibnamefont{Collins}},
  \bibinfo{author}{\bibfnamefont{D.~E.} \bibnamefont{Soper}}, \bibnamefont{and}
  \bibinfo{author}{\bibfnamefont{G.}~\bibnamefont{Sterman}},
  \bibinfo{journal}{Nucl. Phys.} \textbf{\bibinfo{volume}{B261}},
  \bibinfo{pages}{104} (\bibinfo{year}{1985}).

\bibitem[{\citenamefont{Bodwin}(1985)}]{Bodwin:1984hc}
\bibinfo{author}{\bibfnamefont{G.~T.} \bibnamefont{Bodwin}},
  \bibinfo{journal}{Phys. Rev.} \textbf{\bibinfo{volume}{D31}},
  \bibinfo{pages}{2616} (\bibinfo{year}{1985}).

\bibitem[{\citenamefont{Altarelli and Parisi}(1977)}]{Altarelli:1977zs}
\bibinfo{author}{\bibfnamefont{G.}~\bibnamefont{Altarelli}} \bibnamefont{and}
  \bibinfo{author}{\bibfnamefont{G.}~\bibnamefont{Parisi}},
  \bibinfo{journal}{Nucl. Phys.} \textbf{\bibinfo{volume}{B126}},
  \bibinfo{pages}{298} (\bibinfo{year}{1977}).

\bibitem[{\citenamefont{Altarelli et~al.}(1979)\citenamefont{Altarelli, Ellis,
  and Martinelli}}]{Altarelli:1979ub}
\bibinfo{author}{\bibfnamefont{G.}~\bibnamefont{Altarelli}},
  \bibinfo{author}{\bibfnamefont{R.~K.} \bibnamefont{Ellis}}, \bibnamefont{and}
  \bibinfo{author}{\bibfnamefont{G.}~\bibnamefont{Martinelli}},
  \bibinfo{journal}{Nucl. Phys.} \textbf{\bibinfo{volume}{B157}},
  \bibinfo{pages}{461} (\bibinfo{year}{1979}).


\bibitem{cteq}
J. Pumplin, D.R. Stump, J.Huston, H.L. Lai, P. Nadolsky and W.K. Tung, JHEP 0207:012(2002).

\bibitem[{\citenamefont{Abulencia et~al.}(2005)}]{Abulencia:2005us}
\bibinfo{author}{\bibfnamefont{A.}~\bibnamefont{Abulencia}}
  \bibnamefont{et~al.} (\bibinfo{collaboration}{CDF}), 
  \bibinfo{journal}{Phys. Rev.} \textbf{\bibinfo{volume}{D71}},
  \bibinfo{pages}{032001} (\bibinfo{year}{2005}).

\bibitem{Acosta:2001gv}
  D.~E.~Acosta {\it et al.}  [CDF Collaboration],
  Phys.\ Rev.\ Lett.\  {\bf 88}, 161802 (2002).

\end{thebibliography}
\end{document}